\documentclass{article}
\usepackage[utf8]{inputenc}
\usepackage{hyperref}
\usepackage{graphicx}
\usepackage{authblk}
\usepackage{amsmath}
\usepackage{xcolor} 
\usepackage{soul} 

\usepackage[natbibapa,nodoi]{apacite}

\usepackage[margin=0.984252in]{geometry}

\usepackage{setspace}

\title{Unveiling the Impact of Social and Environmental Determinants of Health on Lung Function Decline in Cystic Fibrosis through Data Integration using the US Registry}

\author[1, 2, 3]{Eleni-Rosalina Andrinopoulou}\author[4, 5]{Emrah Gecili}\author[4, 5, 6]{Rhonda D Szczesniak}
\affil[1]{Department of Biostatistics, Erasmus Medical Center, the Netherlands}
\affil[2]{Department of Epidemiology, Erasmus Medical Center, the Netherlands}
\affil[3]{Statistics and Data Science Innovation Hub, GSK, the Netherlands}
\affil[4]{Division of Biostatistics \& Epidemiology, Cincinnati Children’s Hospital Medical Center, Cincinnati, OH, USA}
\affil[5]{Department of Pediatrics, University of Cincinnati Children’s Hospital Medical Center, Cincinnati, OH, USA}
\affil[6]{Division of Pulmonary Medicine, Cincinnati Children’s Hospital Medical Center, Cincinnati, OH, USA}

\date{}
\begin{document}

\maketitle

\section*{Abstract}

Integrating diverse data sources offers a comprehensive view of patient health and holds potential for improving clinical decision-making. In Cystic Fibrosis (CF), which is a genetic disorder primarily affecting the lungs, biomarkers that track lung function decline such as FEV$_1$ serve as important predictors for assessing disease progression. Prior research has shown that incorporating social and environmental determinants of health improves prognostic accuracy. To investigate the lung function decline among individuals with CF, we integrate data from the U.S. Cystic Fibrosis Foundation Patient Registry with social and environmental health information. Our analysis focuses on the relationship between lung function and the deprivation index, a composite measure of socioeconomic status.

We used advanced multivariate mixed-effects models, which allow for the joint modelling of multiple longitudinal outcomes with flexible functional forms.  This methodology provides an understanding of interrelationships among outcomes,
addressing the complexities of dynamic health data.  We examine whether this relationship varies with patients’ exposure duration to high-deprivation areas, analyzing data across time and within individual US
states. Results show a strong relation between lung function and the area under the deprivation index
curve across all states. These results underscore the importance of integrating social and environmental determinants of health into clinical models of disease progression. By accounting for broader contextual factors, healthcare providers can gain deeper insights into disease trajectories and design more targeted intervention strategies.

\textbf{Keywords}: Cystic Fibrosis, data integration, environmental markers, multivariate mixed models, association structure


\section{Introduction}

Integrating multiple data information in healthcare provides a panoramic view of the patient's care, which can enhance clinical decision-making in the management of chronic disease. The ability to collect a large amount of data has made merging diverse health-related information from various sources popular. Cystic Fibrosis (CF) is a genetic disorder that affects organs such as the lungs, pancreas, and intestines. The respiratory condition represents the main cause of mortality; therefore, lung function decline is one of the primary factors that monitor the progression of the disease \citep{andrinopoulou2020multivariate, liou2001predictive}. CF also leads to pancreatic insufficiency, intestinal motility changes, and excessive enterocyte mucus. Impairment of these organs can result in inadequate growth; thus, the nutritional status of the patients is normally also monitored using Body Mass Index (BMI) percentile \citep{liou2001predictive, stallings2008evidence}. Despite previous examination of the relationship between BMI and FEV$_1$ (Forced Expiratory Volume in one second) through multivariate joint models, which linked these two variables with the risk of acute respiratory events known as pulmonary exacerbations, there has been limited investigation into additional factors that influence the progression of the disease \citep{andrinopoulou2020multivariate}. 

There has been an increasing interest in integrating markers of social and environmental determinants of health with clinical data to investigate further factors that influence the progression of the disease \citep{szczesniak2023lung, collaco2010quantification, taylor2013effect, goeminne2013impact}. Incorporating such information could improve decisions regarding the monitoring and treatment strategies of the patients. Previous work focusing on such data integration illustrates the importance of environmental and community factors on the progression of lung function decline \citep{gecili2023built, palipana2023social}. In particular, greenspace, limited elemental carbon attributable to traffic sources (a marker of traffic related air pollution), and low community deprivation seem to improve the prognosis of FEV$_1$ decline in CF. Several socioeconomic variables (such as poverty, assisted income, median income, education, vacant housing, and healthcare access) from the American Community Survey (ACS)  capture “community deprivation”. Because of the strong correlation among those variables, previous studies concentrated on combining them \citep{brokamp2019material}. They used principal components analysis of six different 2015 ACS measures and the deprivation index marker was defined as the first component, which explains over 60\% of the total variance. The resulting index was rescaled and normalized to range from zero to one, with higher values indicating greater deprivation. Despite the intuitive connection between social and environmental determinants of health and biomarkers in respiratory diseases, there has been insufficient exploration into effectively incorporating those two types of information into a single statistical model. Such a model would assess the extent to which exposure to specific socioeconomic groups and geographic regions affects disease progression.

The deprivation index provides a measure of the environment in which individuals reside, influencing their health outcomes \citep{mayhew2024social, roux2001commentary}. The community deprivation index has been previously used as a proxy for individual deprivation because individual-level socioeconomic data is often unavailable or difficult to obtain. However, considering living in a deprived area as an index of community has several limitations. Not everyone in a deprived area experiences the same level of deprivation. Furthermore, people's socioeconomic status can change, and they might move between areas of varying deprivation levels, which static indices fail to capture. To address these issues, individual residence as a dynamic variable is essential. Implementing longitudinal studies that track individuals over time, capturing their movement and changes in residential environments, allows us to account for these individual variations. This dynamic approach helps understand how changes in social and environmental determinants of health impact biomedical outcomes.

There are multiple ways of investigating the connection between multiple longitudinal outcomes, such as deprivation index and lung function. A straightforward approach is to use one response as a time-dependent covariate, but this method has several limitations \citep{delporte2024joint}. In particular, incorrectly specifying the lag can lead to illogical results. Such lag assumption is also important in the case of social and environmental determinants, where the exposure in the past period can be more influential on the outcome than the current exposure. Furthermore, distinguishing between exogenous and endogenous covariates is crucial. A time-varying covariate is considered endogenous if its value at a particular time point is not conditionally independent of all preceding outcome measurements, meaning it is impacted by prior outcome values. Some environmental factors can be considered endogenous variables because they often interact with other variables, such as, socioeconomic status, assess to care, and psychosocial stress. Finally, handling missing data is challenging. While missing values in the outcome can be handled under the assumption of missing at random using likelihood methods, this approach does not apply to missing values in covariates. When data are collected at irregular intervals, using time-dependent covariates with lags is problematic. An alternative approach to overcome these issues is to use multivariate models. We developed an advanced multivariate mixed-effects model to establish meaningful connections between multiple longitudinal outcomes, incorporating functional forms tailored to reflect biological processes and assumptions. This approach not only aligns with the underlying physiological mechanisms, but also enhances interpretability, offering a clear and intuitive understanding of the dynamic relationships between outcomes. Building on previous work that explored related methodologies in longitudinal data analysis \citep{andrinopoulou2020multivariate, liou2001predictive}, our framework provides a robust basis for uncovering clinically actionable insights from complex datasets.

The relation between commonly used biomarkers that measure lung function decline with social and environmental determinants of health could reveal new insights into disease development and the potential influence of geography. This would further explain how the natural disease progression of CF may be impacted by where an individual patient resides. In this work, using the US Cystic Fibrosis Foundation Patient Registry (CFFPR) data \citep{knapp2016cystic} including over 35,000 CF patients with over 350,000 years of cumulative follow-up, we investigate the relation between FEV$_1$ percentage predicted with the deprivation index. The CFFPR consists of demographics and annual or encounter-based outcomes on individuals living with CF in the US. We further explore whether this connection changes when patients are exposed longer or shorter to high deprivation index areas. Given the dependency of social and environmental determinants on geographic regions, we investigate their relation with FEV$_1$ percentage predicted across all US states. The rest of the paper is organized as follows. Section 2 describes the methods. In Section 3, we present a sensitivity analysis regarding the different association structures, and in Section 4, we fit the proposed methodology on the US CFFPR data. Finally, section 5 concludes with a discussion of the implications and insights derived from the results.

\section{Methods}

\subsection{Data sources}

The CFFPR data collected from 1997 to 2017 was linked to social and environmental determinants of health using ZIP code tabulation areas (ZCTAs, a census-derived geography for 5-digit residential ZIP Codes). Patients older than six years with at least one measurement for each outcome were included in the study. Data observed after lung transplant were censored, as these data represent lung function that is not indicative of natural disease progression.

\subsection{Statistical models}

Connecting different outcomes that are repeatedly measured, such as FEV$_1$ and community deprivation index, can present challenges. Standard approaches for the analysis of such data include the use of multivariate mixed-effects models. We consider a total of $n$ subjects, and we let $y_{i1}(t)$ denote the follow-up measurements for the longitudinal outcome FEV$_1$ measured as percentage predicted for patient $i$, where $i = 1,\dots,n$, at time point $t$. Here, time refers to the age at a given clinical encounter. Similarly, let $y_{i2}(t)$ denote the follow-up measurements for the longitudinal outcome of deprivation index. These measurements may be obtained at various time points and may vary in duration for each patient. We integrated the outcomes of interest by employing a multivariate mixed-effects model to characterize the individualized developments of the longitudinal outcomes over time. In particular, if $k = 1, \dots, K$ denotes the longitudinal outcome, we postulate

\begin{eqnarray*}
y_{ik}(t) &=& \boldsymbol{x}_{k}^\top (t) \boldsymbol{\beta_k} + \boldsymbol{z}_{ik}^\top (t) \boldsymbol{b}_{ik} + \epsilon_{ik}(t) = m_{ik}(t) + \epsilon_{ik}(t),
\end{eqnarray*}
where, $\boldsymbol{x}_{k} (t)$ denote the design vectors for the fixed effects regression coefficients $\boldsymbol{\beta_k}$. Likewise, $\boldsymbol{z}_{ik}(t)$ denote the design vectors for the random effects $\boldsymbol{b}_{ik}$.
In common practice, the two longitudinal outcomes are connected via the random effects where a multivariate normal distribution is assumed $\boldsymbol{b}_{i}=(\boldsymbol{b}_{i1}, \dots, \boldsymbol{b}_{iK}) \sim N(0, \boldsymbol{D})$, where $\boldsymbol{D}$ is the variance-covariance matrix of the random effects.

\begin{figure}[!ht]
    \centerline{%
    \includegraphics[width = 7cm]{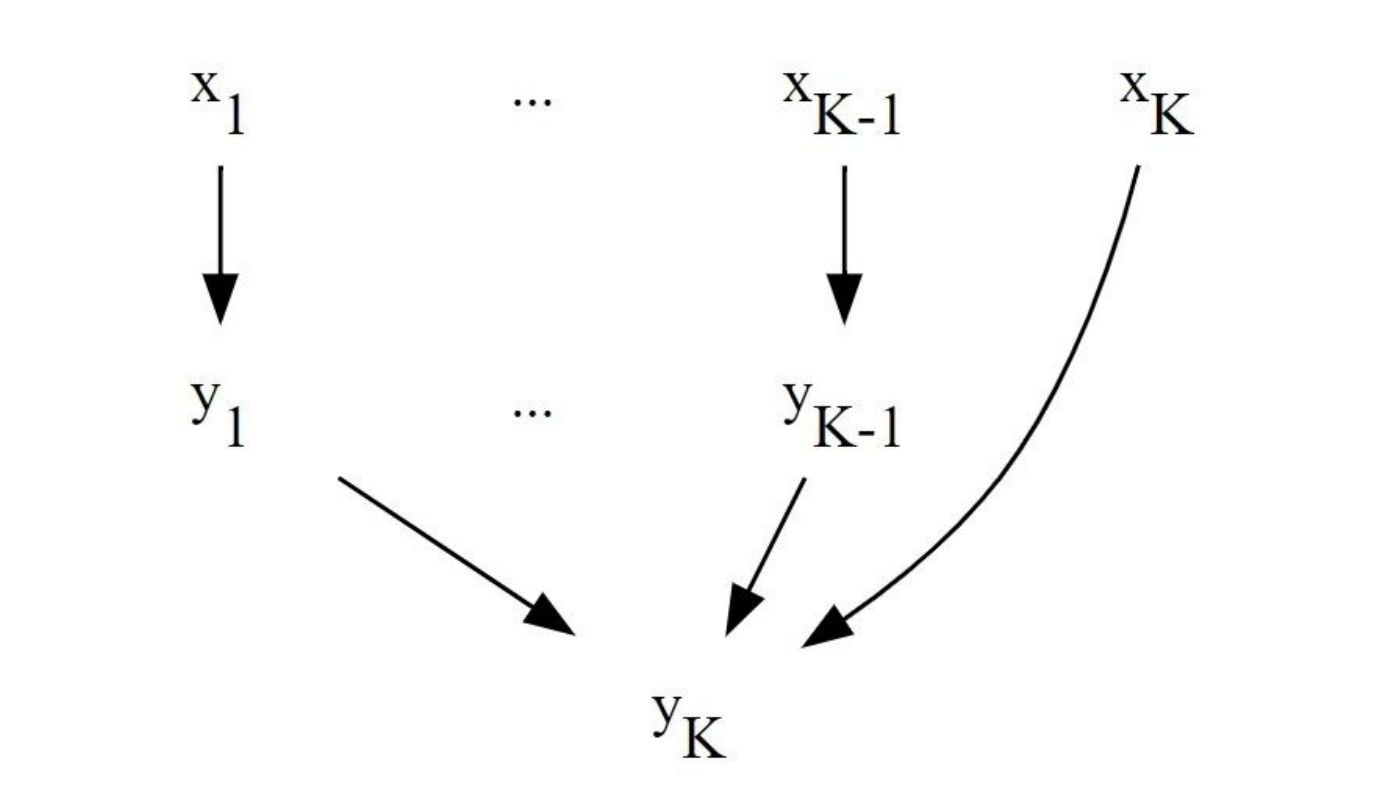}}
    \caption{Graphical representation of associations between covariates and outcomes. Outcomes are presented as y and covariates are presented as x.}
    \label{fig:DAG_gen}
\end{figure}

A challenge in such models lies in interpreting the parameters that represent the association, which in the aforementioned case are the parameters of the variance-covariance matrix of the random effects. These parameters do not directly measure the strength of the association and lack clinical relevance.  Integrating longitudinal outcomes as time-dependent covariates into the primary model would provide a clearer understanding of the association parameters. An example of how the different types of information could be connected is illustrated in Figure \ref{fig:DAG_gen}. To assess the strength of the connection between social and environmental determinants of health with biomarkers, we propose to add an association parameter that relates the two longitudinal outcomes. In particular, we integrate the underlying value of the outcome $m_{ik}(t)$ ($k \neq K$), e.g., the deprivation index, into the model of the primary outcome, e.g., the FEV$_1$ outcome. Even though the deprivation index is a time-dependent variable, significant fluctuations are not anticipated, and the underlying value observed at the same time as the lung function would have minimal effect on the progression of the disease. Therefore, we assume different functional forms, such as the deprivation index's cumulative effect (area under the curve), to be connected with lung function at a specific time point. We illustrate this relationship in the Appendix Figure S1. The multivariate mixed-effects model can then be extended as shown below:

\begin{eqnarray*}
y_{iq}(t) &=& \boldsymbol{x}_{iq}^\top (t) \boldsymbol{\beta_q} + \boldsymbol{z}_{iq}^\top (t) \boldsymbol{b}_{iq} + \epsilon_{iq}(t) = m_{iq}(t) + \epsilon_{iq}(t),\\
y_{iK}(t) &=& \boldsymbol{x}_{iK}^\top (t) \boldsymbol{\beta_K} + \boldsymbol{z}_{iK}^\top (t) \boldsymbol{b}_{iK} + \epsilon_{iK}(t) = m_{iK}(t) + \alpha \frac{1}{t}\int^t_0 m_{iq}(s)ds + \epsilon_{iK}(t),
\label{MM2a}
\end{eqnarray*}

where $q = 1, \dots, K-1$ represents all outcomes apart from the outcome of primary interest. The association parameter $\alpha$ indicates the relationship between the markers. The term $\frac{1}{t}$ would normalize the area under the curve to overcome convergence issues. The interpretation of the association parameter is straightforward, and it presents the one unit difference in the outcome $y_{i1}(t)$ when the normalized area under the curve of $y_{i2}(t)$ changes by 1 unit, given that all other covariates remain constant. Since the length of exposure would have a different effect on the progression of the disease, a further extension includes using different periods for the cumulative effect. In particular, we can focus on $d = 2, 5, 10, 15$ years before a particular time point. In the Appendix, Figure S2 shows an example when assuming two years before the FEV$_1$ percentage predicted measurement. Then the model takes the form:

\begin{eqnarray*}
y_{iq}(t) &=& \boldsymbol{x}_{iq}^\top (t) \boldsymbol{\beta_q} + \boldsymbol{z}_{iq}^\top (t) \boldsymbol{b}_{iq} + \epsilon_{iq}(t) = m_{iq}(t) + \epsilon_{iq}(t),\\
y_{iK}(t) &=& \boldsymbol{x}_{iK}^\top (t) \boldsymbol{\beta_K} + \boldsymbol{z}_{iK}^\top (t) \boldsymbol{b}_{iK} + \epsilon_{iK}(t) = m_{iK}(t) + \alpha\frac{1}{t}\int^t_{t-d} m_{iq}(s)ds + \epsilon_{i1}(t)
\label{MM2b}
\end{eqnarray*}

\subsection{Estimation}
We assume the Bayesian framework, and specifically, we implemented a Markov Chain Monte Carlo (MCMC) algorithm to obtain the parameters of interest by utilizing the corresponding posterior distributions and the data. Software to implement the proposed methodology is publicly available on: \newline
https://github.com/ERandrinopoulou/multiLME. 

The likelihood of the model is derived under the assumption that the longitudinal processes are independent given the random effects. Moreover, the
longitudinal responses of each subject are assumed independent given the random effects. When no prior knowledge is available, non-informative priors for all the parameters can be assumed. In particular, we use a Normal distribution with mean zero and variance 100 for the fixed effects coefficients ($\beta_1$ and $\beta_2$) and the association parameter ($\alpha$). For the precision of the error terms ($\epsilon_{i1}$ and $\epsilon_{i2}$), a Gamma distribution is assumed with both shape and rate parameters set to 0.01. For the random effects precision matrix, a Wishart distribution is assigned with degrees of freedom equal to the total number of random effects plus one and the scale matrix is determined by a hyperprior Gamma distribution (4* \texttt{Gamma}(0.5, 0.01)). We employed two MCMC chains, each with 28,000 iterations, including a burn-in period of 3,000 iterations and a thinning interval of 50. For the adaptive phase, we assume 3000 iterations. Trace plots and the Gelman–Rubin diagnostic were used to evaluate the MCMC convergence \citep{gelman1992inference}. All analyses were implemented using R (version 4.5.0). 

\section{Sensitivity analysis}

A challenge in complex models, such as multivariate mixed-effect models, that integrate different sources of information is the choice of the most appropriate association structure. As illustrated previously, a connection between the different longitudinal outcomes could be established in several ways, e.g., via the random effects or an association parameter. It is not always straightforward whether both structures should be assumed in each application and in some cases, connecting the outcomes in multiple ways could lead to computational problems. Therefore, we performed a sensitivity analysis to investigate the amount of bias introduced when we ignore or add an extra way to connect the different data sources. Checking the relationship between the outcomes in a real-world data set is a difficult task, therefore we performed a simulation study where the true association structure is known. We conducted simulations where an association parameter links the outcomes. Specifically, this involves incorporating the second outcome into the model for the first outcome. Subsequently, we applied models that assume connections between outcomes through an association parameter, random effects, or both. We denote these three distinct structures as value/slope/area, RE, and value/slope/area \& RE, respectively. In the Appendix, Figures S3-S5 illustrate the results when simulating a value and a slope association structure, respectively. We obtained, especially in the value parametrization, bias in the parameters of the fixed effects of the first models and in the association parameter.

\section{Cystic Fibrosis Patient Registry Cohort Application  }
The analysis cohort for this application included 16,811 CF patients from across the US, aged 6-20 years, with more than 340,000 lung function measurements observed 
from 1997 to 2017, after merging the registry and ZIP code data sets and excluding any records with missing value in patients characteristics, lung function or social and environmental determinants of health. These patients, born between 1988 and 2009, have been monitored over time, with the number of measurements per patient ranging from 1 to 184 (median 14). Figures S6 and S7 present data characteristics by state. We observed more patients in Texas and California, whereas Ohio, Alabama, and Washington have the longest median number of visits. Additionally, more female patients are observed in Oregon, and the western US has lower percentage of F508del homozygous patients. Furthermore, Alabama has fewer patients who receive pancreatic enzymes. 

\begin{figure}[!ht]
    \centerline{%
    \includegraphics[width = 12cm]{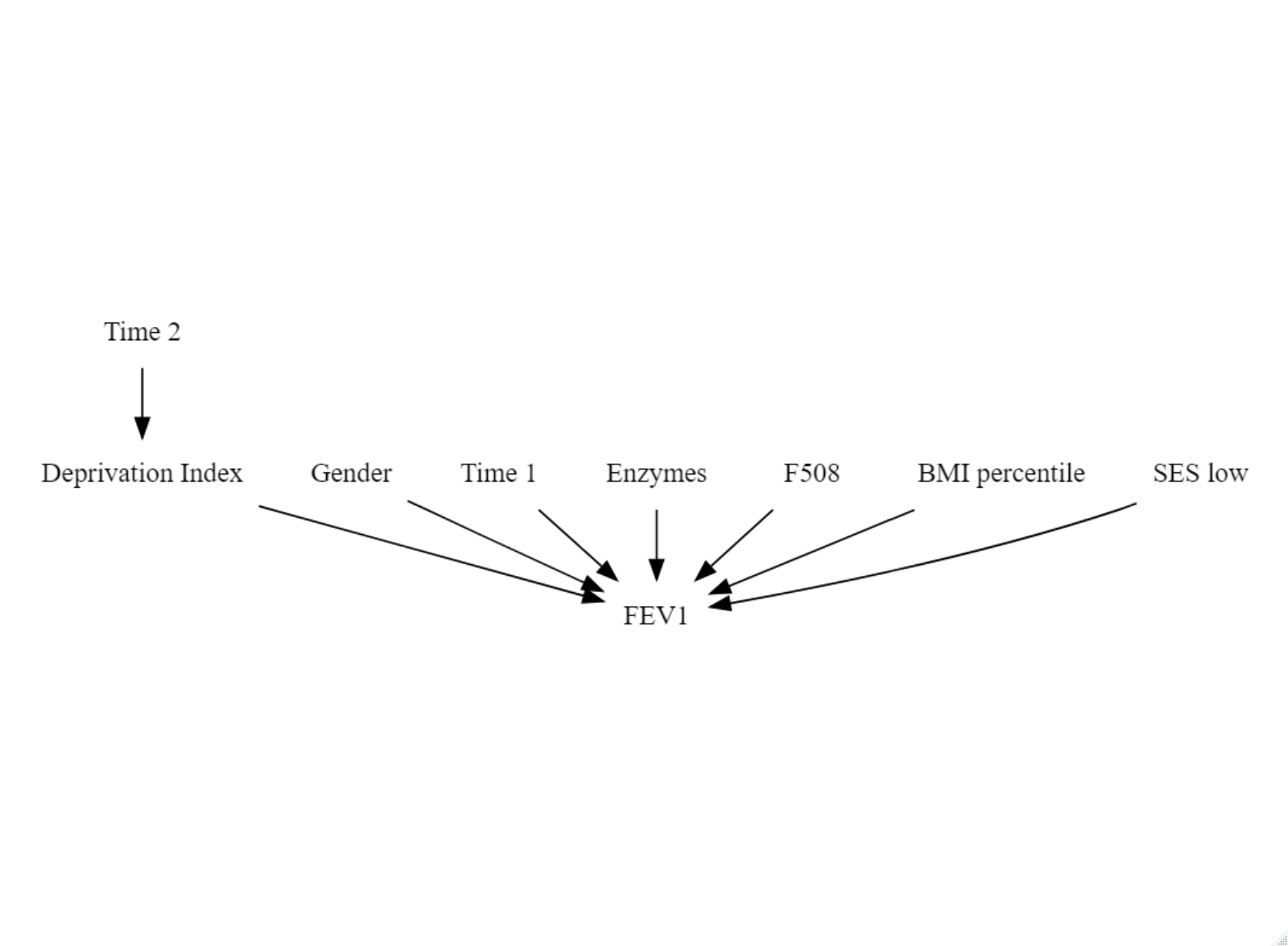}}
    \caption{Example of associations between different information for our CFFPR data application.}
    \label{fig:DAG_spec}
\end{figure}

To investigate the association between FEV$_1$ and the deprivation index, we fitted a multivariate mixed-effects model for each state in the US. An illustration of these associations is presented in Figure \ref{fig:DAG_spec}. For the fixed effects in the FEV$_1$ percentage predicted model, we assumed as covariates: gender, whether the patient comes from a low socioeconomic status, genotype (F508del heterozygous, F508del homozygous, or other), BMI percentile, and whether the patient is on pancreatic enzymes. The CDC growth charts were used to obtain the missing BMI percentile after the age of 18 \citep{kuczmarski20022000}. For the calculation, we assumed age at the last BMI percentile and the actual weight and height of the patient. For the time structure, we assumed natural cubic splines with two degrees of freedom for both the fixed and random effects to capture the nonlinear evolution over time (illustrated Appendix Figure S8). In the deprivation index model, we only include time as a covariate with a nonlinear structure (illustrated in Appendix Figure S9). The outcomes FEV$_1$ percentage predicted and deprivation index are measured at the same time points, but the proposed methodology would also work when the time points are different per outcome. The models are illustrated below:

\begin{align*}
\texttt{Deprivation\_Index}_i(t)
&= (\beta_{20}+b_{i20})
  + \sum_{u=1}^2 (\beta_{2u}+b_{i2u})\, ns(t)
  + \epsilon_{i2}(t) \\
\texttt{FEV\_\%\_predicted}_i(t)
&= (\beta_{10}+b_{i10})
  + \sum_{u=1}^2 (\beta_{1u}+b_{i1u})\, ns(t)
  + \beta_{13} \texttt{Gender}_i
  + \beta_{14} \texttt{SESlow}_i \\[0.5em]
&\quad
  + \sum_{u=5}^6 \beta_{1u} \texttt{F508}_i
  + \beta_{17} \texttt{BMIpercentile}_i
  + \beta_{18} \texttt{Enzymes}_i \\[0.5em]
&\quad
  + \alpha\, \texttt{AUC}(t-d)\,
    \texttt{Deprivation\_Index}_i
  + \epsilon_{i1}(t).
\end{align*}

where $d  = 0, 2, 5, 10, 15$ and $ns(t)$ denotes the natural splines. 

\begin{figure}[!ht]
    \centerline{%
    \includegraphics[width = 15cm]{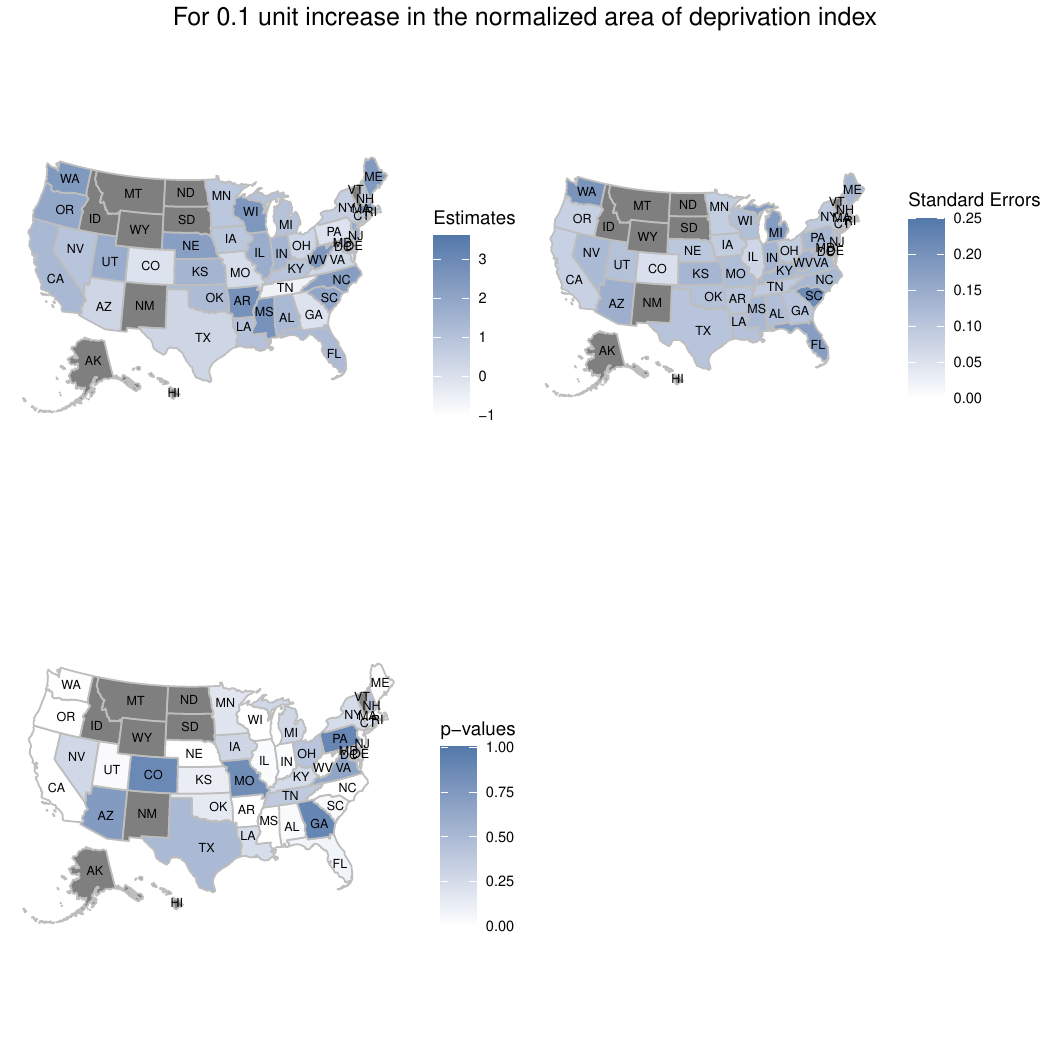}}
    \caption{Association parameters per state. Grey areas indicate the stages where the analysis was not performed due to the small number of patients (n $<$ 120). All longitudinal measurements were used. }
    \label{fig:res_all}
\end{figure}

\begin{figure}[!ht]
    \centerline{%
    \includegraphics[width = 15cm]{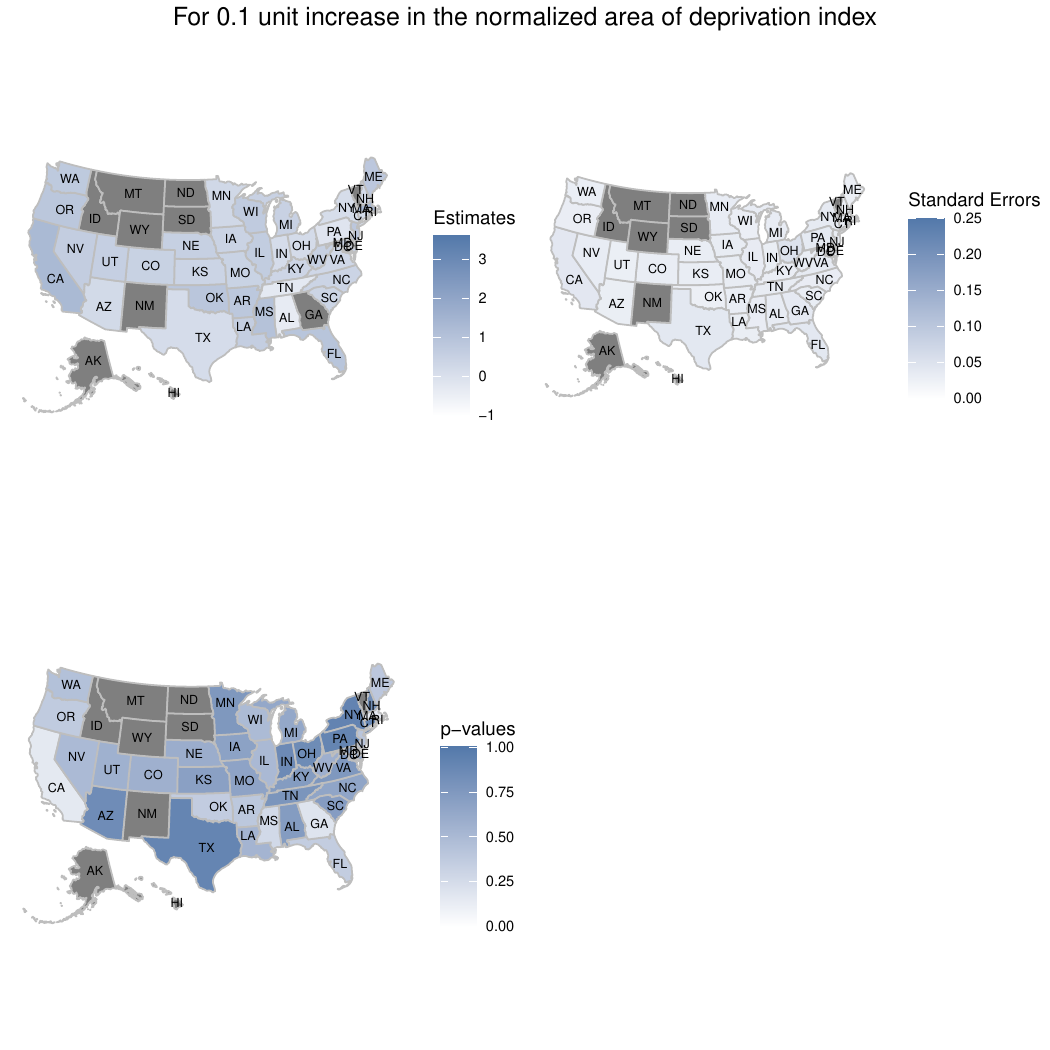}}
    \caption{Association parameters per state. Grey areas indicate the stages where the analysis was not performed due to the small number of patients (n $<$ 120). Two years of longitudinal measurements were used.}
    \label{fig:res_2years}
\end{figure}

\begin{figure}[!ht]
    \centerline{%
    \includegraphics[width = 15cm]{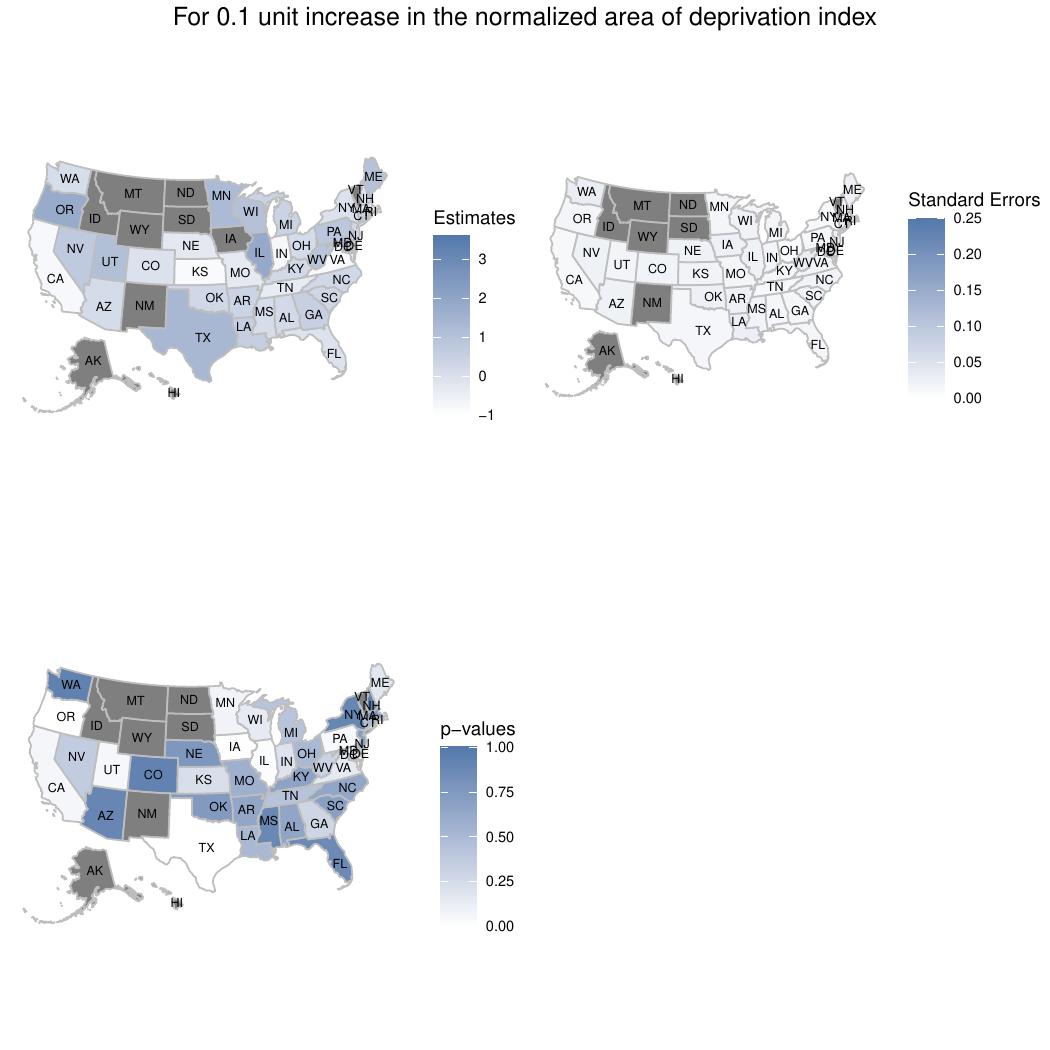}}
    \caption{Association parameters per state. Grey areas indicate the stages where the analysis was not performed due to the small number of patients (n $<$ 120). Five years of longitudinal measurements were used.}
    \label{fig:res_5years}
\end{figure}

\begin{figure}[!ht]
    \centerline{%
    \includegraphics[width = 15cm]{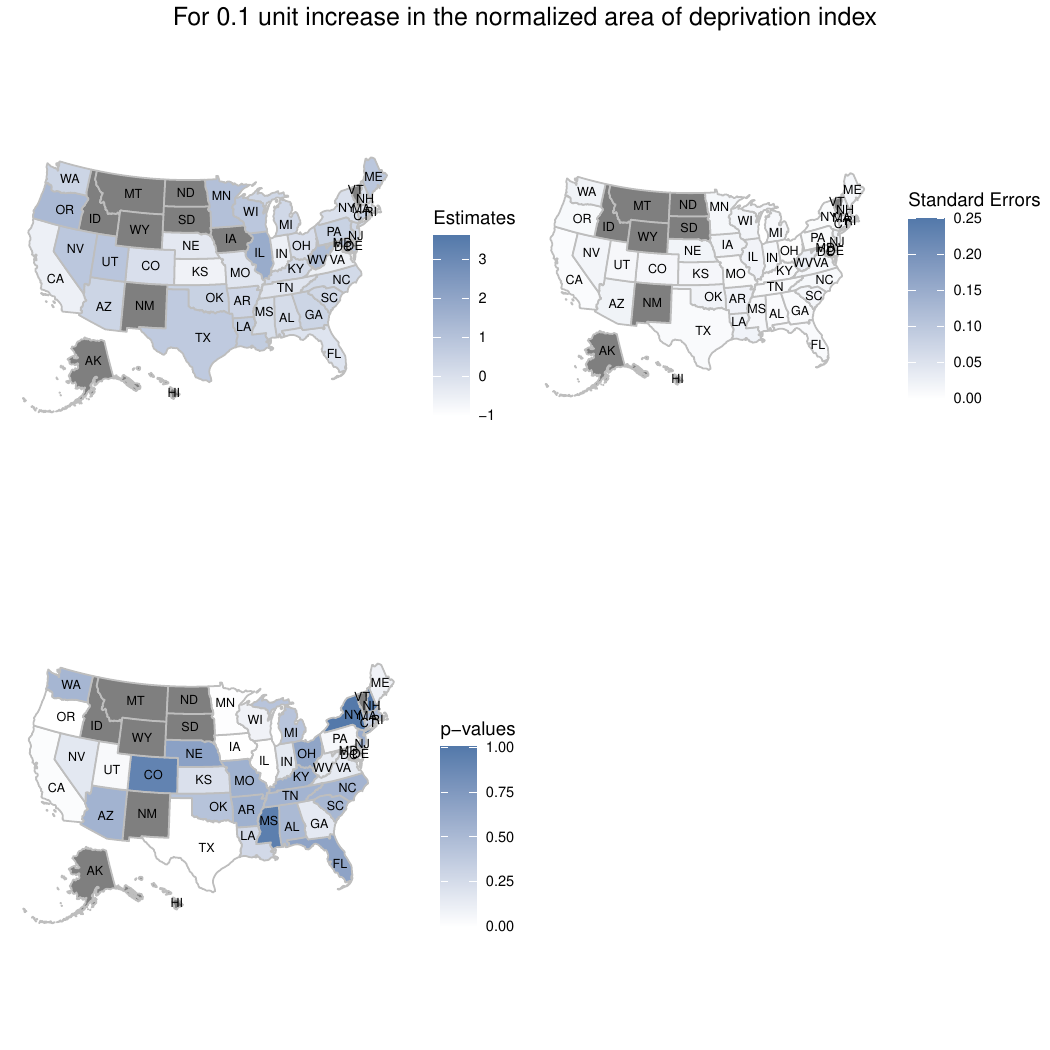}}
    \caption{Association parameters per state. Grey areas indicate the stages where the analysis was not performed due to the small number of patients (n $<$ 120). Ten years of longitudinal measurements were used.}
    \label{fig:res_10years}
\end{figure}

\begin{figure}[!ht]
    \centerline{%
    \includegraphics[width = 15cm]{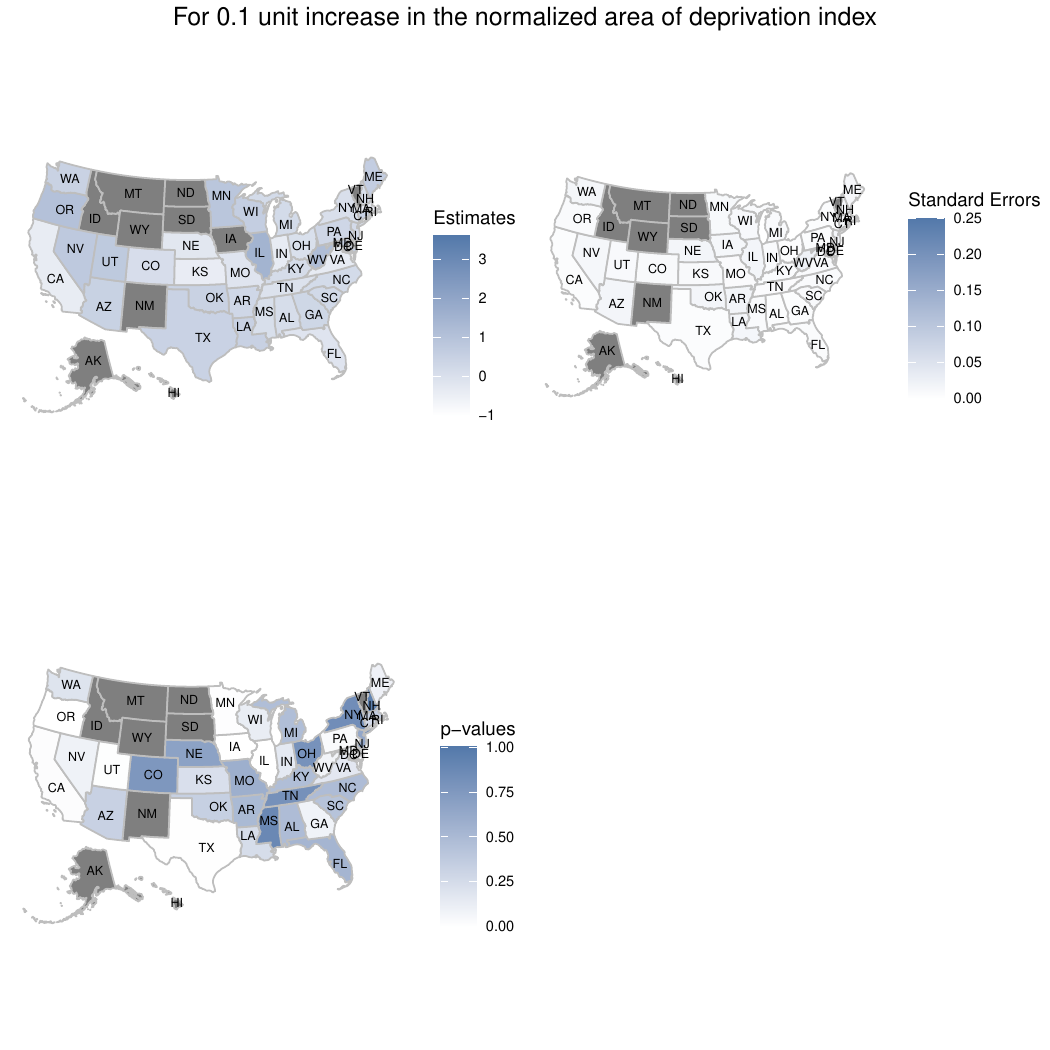}}
    \caption{Association parameters per state. Grey areas indicate the stages where the analysis was not performed due to the small number of patients (n $<$ 120). Fifteen years of longitudinal measurements were used.}
    \label{fig:res_15years}
\end{figure}

To focus on interpreting the association between the outcomes, we assume independence among the random effects. Figures \ref{fig:res_all}- 
\ref{fig:res_15years} present the association parameter $\alpha$ with its standard error and p-value \citep{held2004simultaneous, lesaffre2012bayesian}. To facilitate interpretation and account for the scale of the deprivation index, we transformed the data and focused on a 0.1 unit change. For some regions, the number of patients was insufficient to fit such complex models, which are represented as grey areas in the map ($n$ $<$ 120). We observe that when the whole exposure history of the deprivation index is used or up to 10 years before, the connection between FEV$_1$ with the deprivation index is strong. The strength of this relation remains the same for deprivation measurements used up to 10 and 15 years prior to the FEV$_1$ measurements. We, furthermore, notice that the western and eastern coasts have similar relationships. However, when all measurements are used, the east coast had stronger connections between the outcomes. When the exposure time is short (e.g., 2 or 5 years), the relationship becomes weaker with some differences between the states. Stronger relations between the outcomes are observed in Oregon, Illinois, Minnesota, Wisconsin, Maine and Texas when using measurements up to 5 years.

To investigate whether the relationship between biomarkers and markers of social and environmental determinants of health is the same in different age groups, we repeated the analysis subsetting the data to age at diagnosis as below or equal to 18 and above 12. The results are presented in the Appendix (Figures S10-S19).
Some differences in the results are obtained between the age groups. In particular, stronger connections are observed on the east coast when patients at diagnosis are above 12 years old and all measurements are used. When 2 and 5 years are assumed, there is a stronger connections when the diagnosis age is below 18. In contrast, when 10 and 15 years are assumed, there is a stronger connection when the diagnosis age is above 12.

\section{Discussion}

In this research, we fit a multivariate mixed-effects model to investigate the association of FEV$_1$ with a composite marker of deprivation index. In particular, we included the history of the deprivation index assuming different periods that show how varying extent of exposure may impact lung function. The proposed model was fitted per US state, where drastic differences were observed. In general, the longer a patient is exposed, the stronger the association with FEV$_1$ becomes. Having a large patient registry allowed us to effectively combine complex information and investigate different and more realistic associations, compared to how these associations have been previously studied \citep{gecili2023built}. We found that the effect of community deprivation on lung function is crucial and that the duration of this exposure has an impact on lung function. Furthermore, given the different results per geographical region, environmental factors play an important role in the progression of the CF disease. The results of our research could help identify whether the socioeconomic condition of the patient and neighbourhood influence the disease progression. This could also be the first step towards establishing the "ideal" place to live for a CF patient.   

This study has certain limitations. In particular, the high dimensional data allows more information to be incorporated into the submodels to explore further factors that might impact the associations. A comprehensive indexing of social and environmental determinants of health has revealed that community deprivation is a significant factor contributing to the severe lung function decline phenotype \citep{gecili2023built}. Association studies of this nature require careful assessment with causal inference and control of confounding, but the study presented here provides a blueprint for undertaking this research using multivariate linear mixed-effects models. Considering random effects as independent simplifies certain modeling assumptions. However, a multivariate model remains valuable for capturing dependencies and interactions between different variables and outcomes. It allows us to explore simultaneous relationships and account for correlations among various factors and responses over time. This approach can provide a more comprehensive understanding of the complex dynamics within our data. Physical addresses, which would allow for exact geocoding, were not collected in the motivating dataset, which is typical for many patient registries. Instead, five-digit ZIP codes are readily available and commonly used in health research to approximate local variations in social and environmental determinants that may impact health outcomes \citep{krieger2003race}. However, we acknowledge that the appropriateness of ZIP codes can vary depending on the specific research question. In our study, we recognize that ZIP codes vary in size and may not capture all nuances of smaller neighbourhoods. This choice could potentially influence the spatial granularity of our findings and their generalizability to other geographic regions or populations. 

Future research includes investigating other markers of social and environmental determinants of health disease progression. We focused mainly on the aforementioned marker because we observed changes that the mixed-effects models could capture over time. Other important social and environmental determinants of health seem to be more stable throughout the years \citep{gecili2023built}. Temperature and, in general, climate markers are expected to vary over time and could therefore be incorporated into our model. In our application, we stratified by state, which we expect to capture some climate information. Alternatively, we could include these markers and not stratify by state. A further extension of the model would consist of a single multivariate mixed-effects model that accounts for clustering \citep{palipana2023social}. In particular, the model could be extended to include state as a clustered random effect. Alternatively, spatial covariance extensions could also account for the distances. Further research is also necessary to determine which environmental and socioeconomic factors are more acutely affecting the progression of CF disease. Finally, our methodology could be extended to predict the risk of pulmonary exacerbation, lung transplantation, and mortality outcomes in different environmental characteristics and regions.

\clearpage

\section*{Acknowledgments}

\subsection*{Availability of data and materials}
The authors thank the Cystic Fibrosis Foundation for use of Patient Registry (CFFPR) data to conduct this study and the patients, care providers, and clinic coordinators at U.S. CF centers for their contributions to the CFFPR. Requests to access the CFFPR analysis data utilized for this study may be sent to the Foundation for review/approval: datarequests@cff.org

\subsection*{Funding}
This work was supported by a grant from the National Institutes of Health (R01 HL141286). The content is solely the responsibility of the authors and does not necessarily represent the official views of the National Institutes of Health.

\clearpage

\bibliographystyle{apacite}
\bibliography{references}

\clearpage

\section*{Appendix}

\setcounter{figure}{0}
\renewcommand{\thefigure}{S\arabic{figure}}

\begin{figure}[!ht]
    \centerline{%
    \includegraphics[width = 12cm]{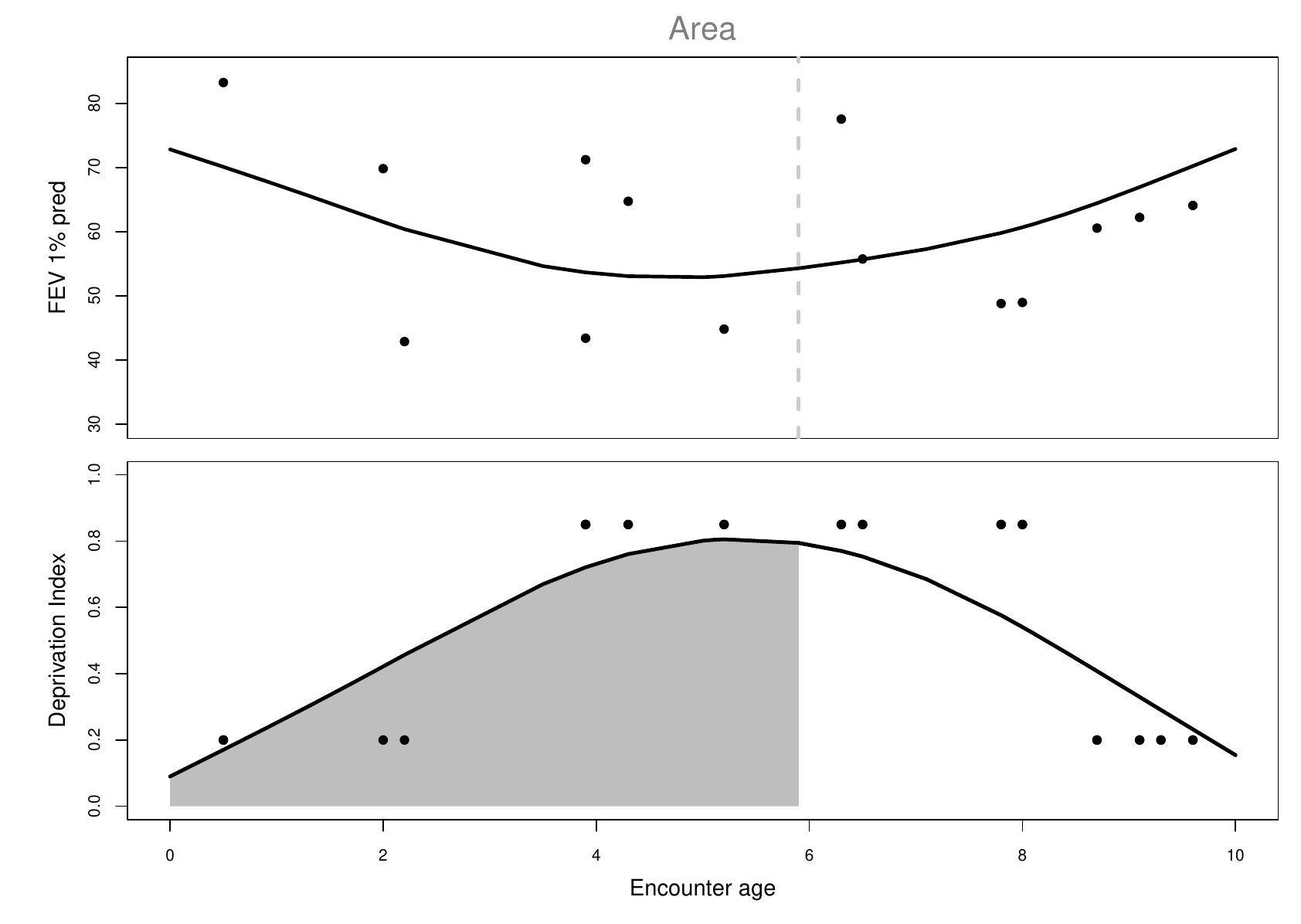}}
    \caption{Graphical representation of the association between the area under the curve of the deprivation index (bottom graph) with the lung function (top graph) at the time point 5.9. In that case, we assume the whole history of the deprivation index.}
    \label{fig:AUC1}
\end{figure}

\begin{figure}[!ht]
    \centerline{%
    \includegraphics[width = 12cm]{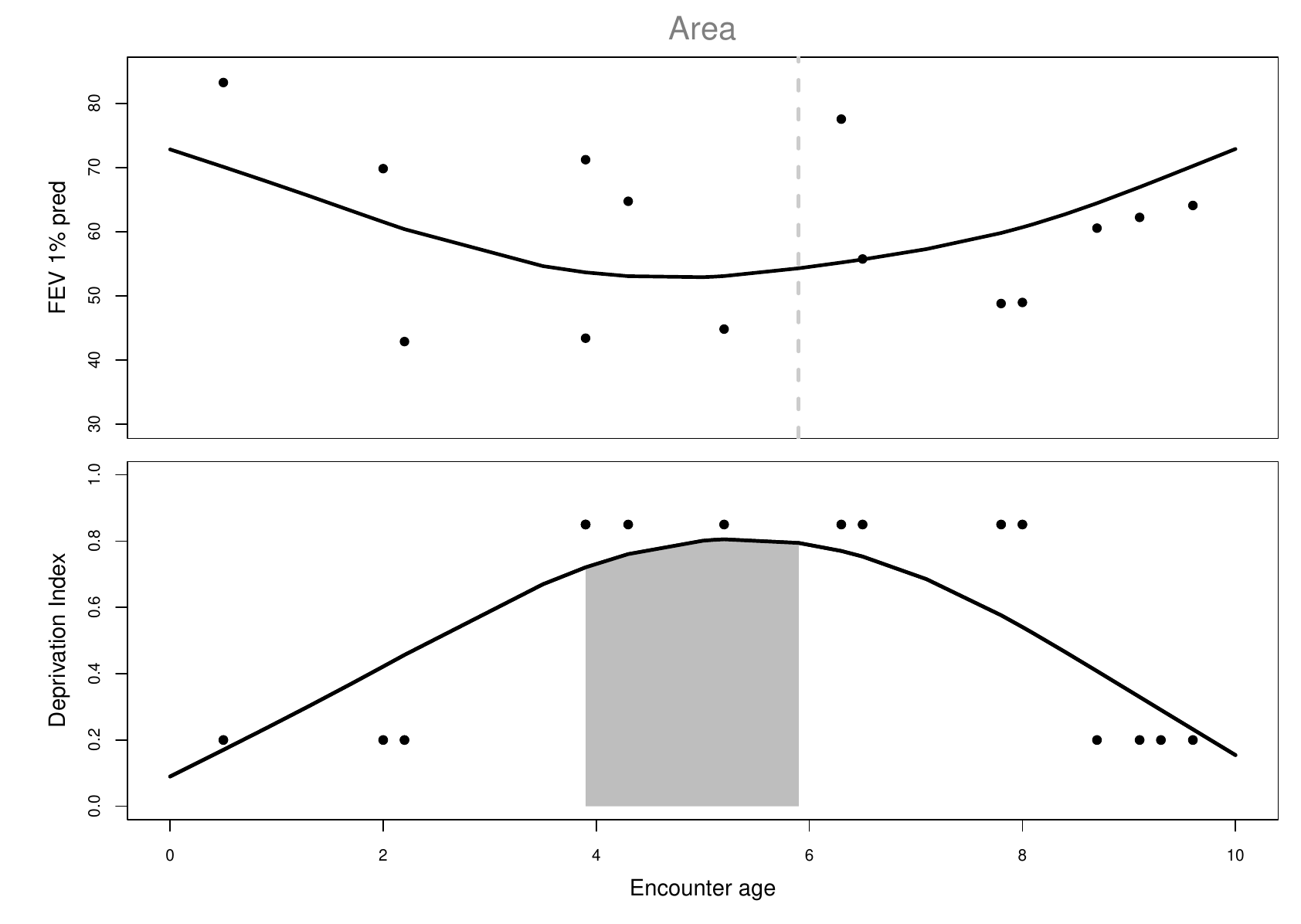}}
    \caption{Graphical representation of the association between the area under the curve of the deprivation index with the lung function process. In that case, we assume a particular time window of the deprivation index (2 years prior to the $FEV_1$ measurement).}
    \label{fig:AUC2}
\end{figure}

\begin{figure}[!ht]
    \centerline{%
    \includegraphics[width = 12cm]{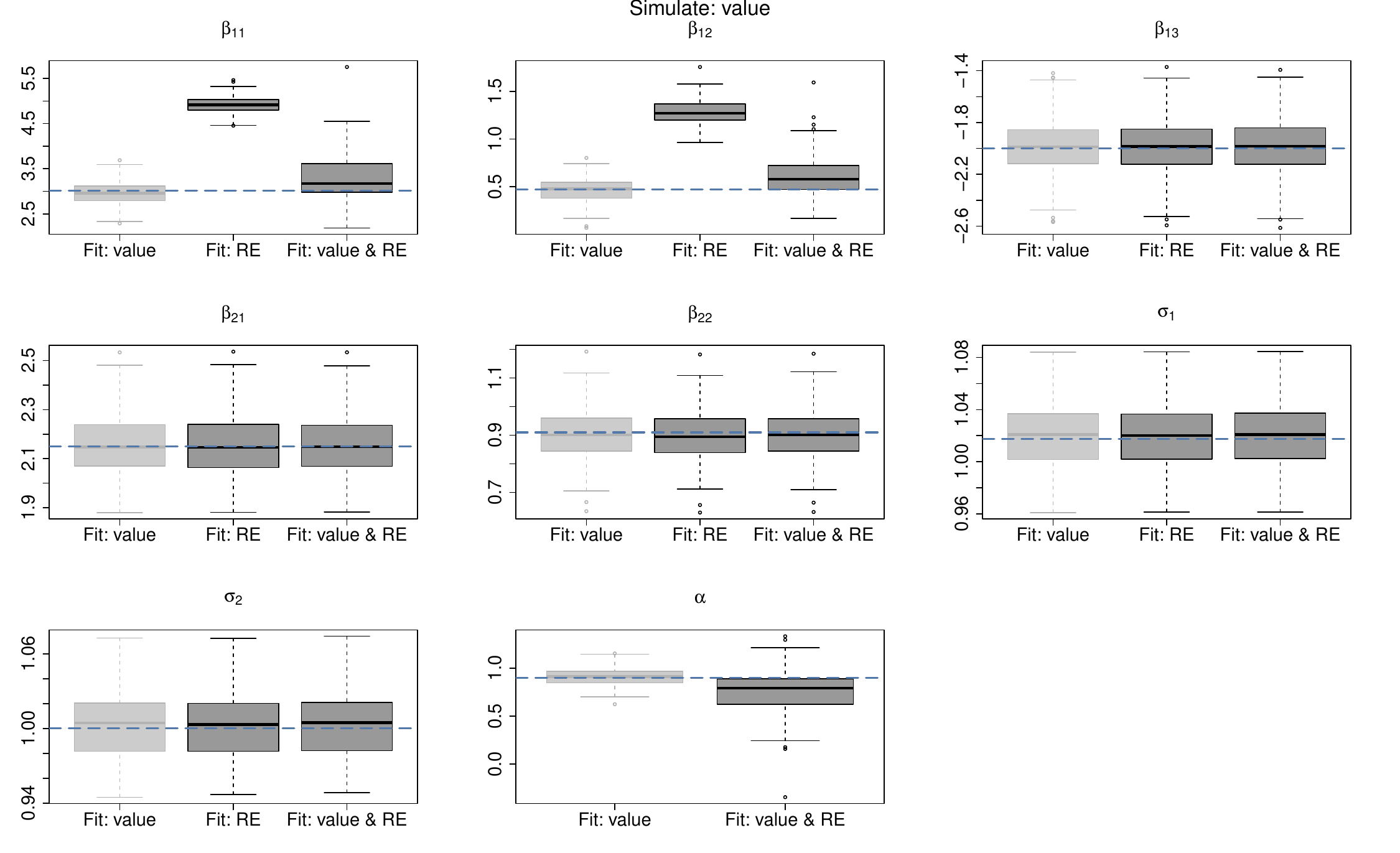}}
    \caption{Simulation results for the value parameterization. Each Boxplot represents the posterior distribution of the parameter estimates from the proposed model for different simulation settings. The graphs are for the estimates of the fixed effects ($\beta_{11}, \beta_{12}, \beta_{13}, \beta_{21}$, and $\beta_{22}$), variance components of random effects ($\sigma_{1}$ and $\sigma_{2}$), and association parameter ($\alpha$) across simulated data sets and model settings. The dashed line indicates the true simulated value for each parameter.}
    \label{fig:Sim_res_value}
\end{figure}

\begin{figure}[!ht]
    \centerline{%
    \includegraphics[width = 12cm]{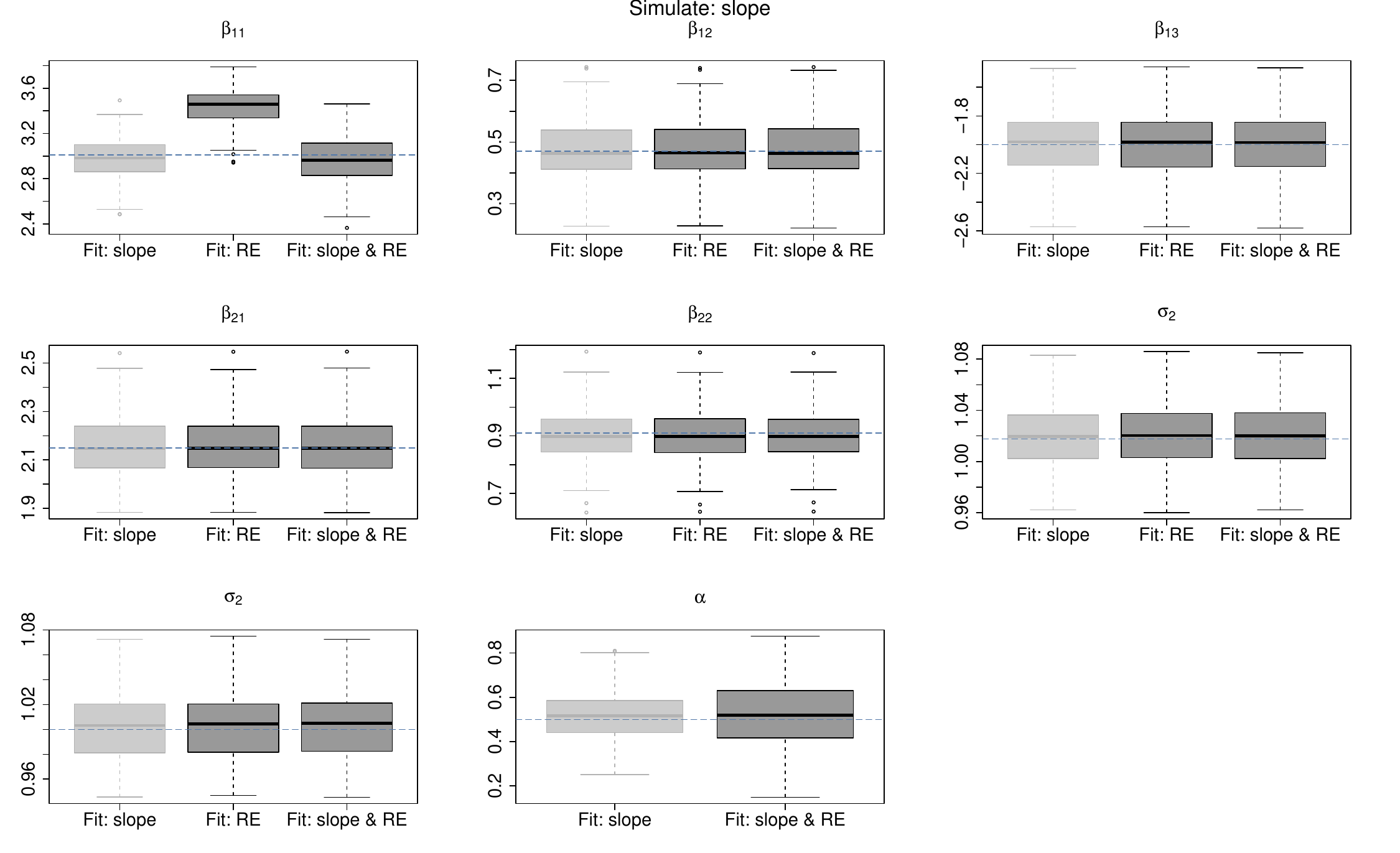}}
    \caption{Simulation results for the slope parameterization. Each Boxplot represents the posterior distribution of the parameter estimates from the proposed model for different simulation settings. The graphs are for the estimates of the fixed effects ($\beta_{11}, \beta_{12}, \beta_{13}, \beta_{21}$, and $\beta_{22}$) , variance components of random effects ($\sigma_{1}$ and $\sigma_{2}$), and association parameter ($\alpha$) across simulated data sets and model settings. The dashed line indicates the true simulated value for each parameter.}
    \label{fig:Sim_res_slope}
\end{figure}

\begin{figure}[!ht]
    \centerline{%
    \includegraphics[width = 12cm]{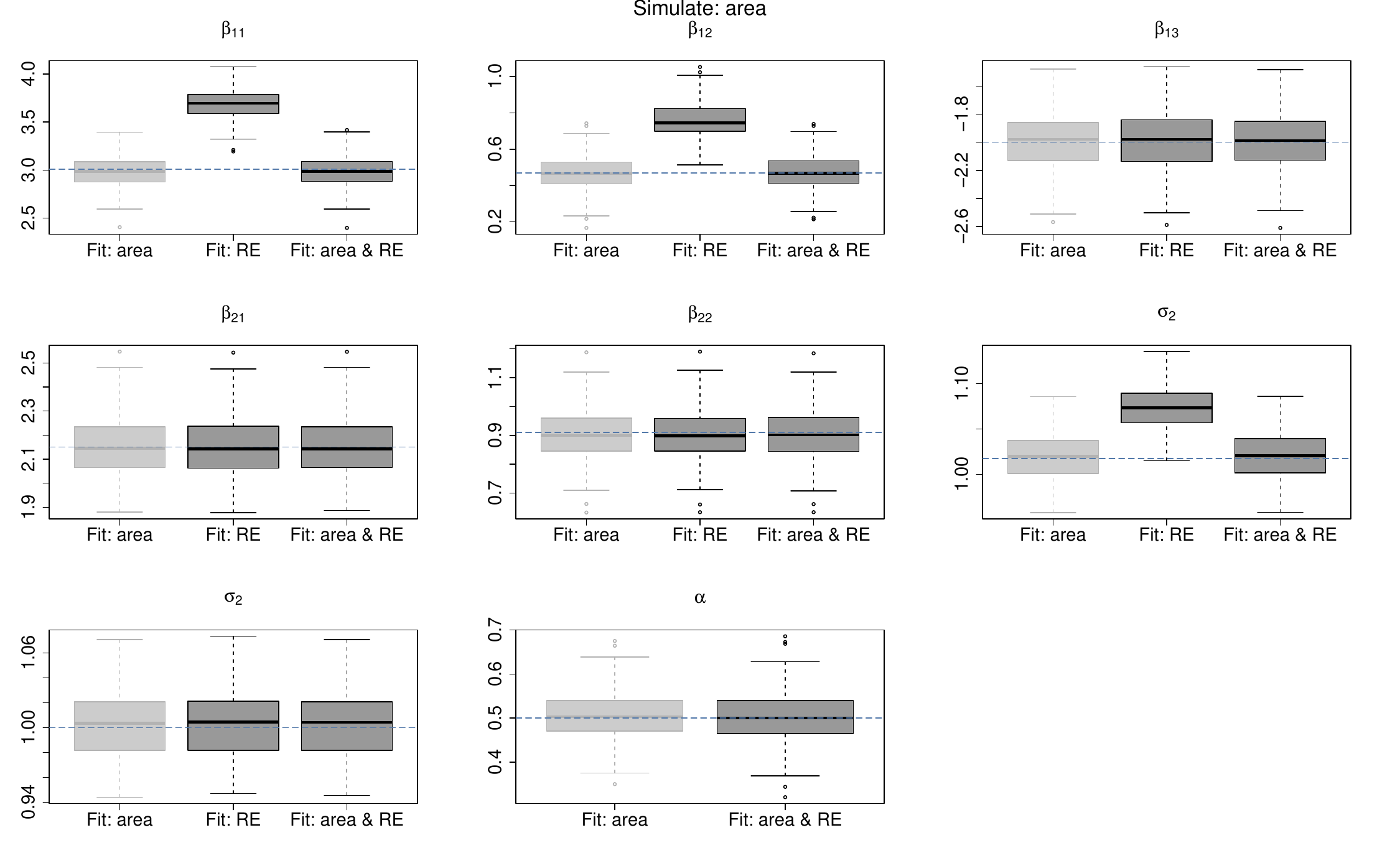}}
    \caption{Simulation results for the area parameterization. Each Boxplot represents the posterior distribution of parameter estimates from the proposed model for different simulation settings. The graphs are for the estimates of fixed effects ($\beta_{11}, \beta_{12}, \beta_{13}, \beta_{21}, and \beta_{22}$) , variance components of random effects ($\sigma_{1}$ and $\sigma_{2}$), and association parameter ($\alpha$) across simulated data sets and model settings. The dashed line indicates the true simulated value for each parameter.}
    \label{fig:Sim_res_area}
\end{figure}

\begin{figure}[!ht]
    \centerline{%
    \includegraphics[width = 15cm]{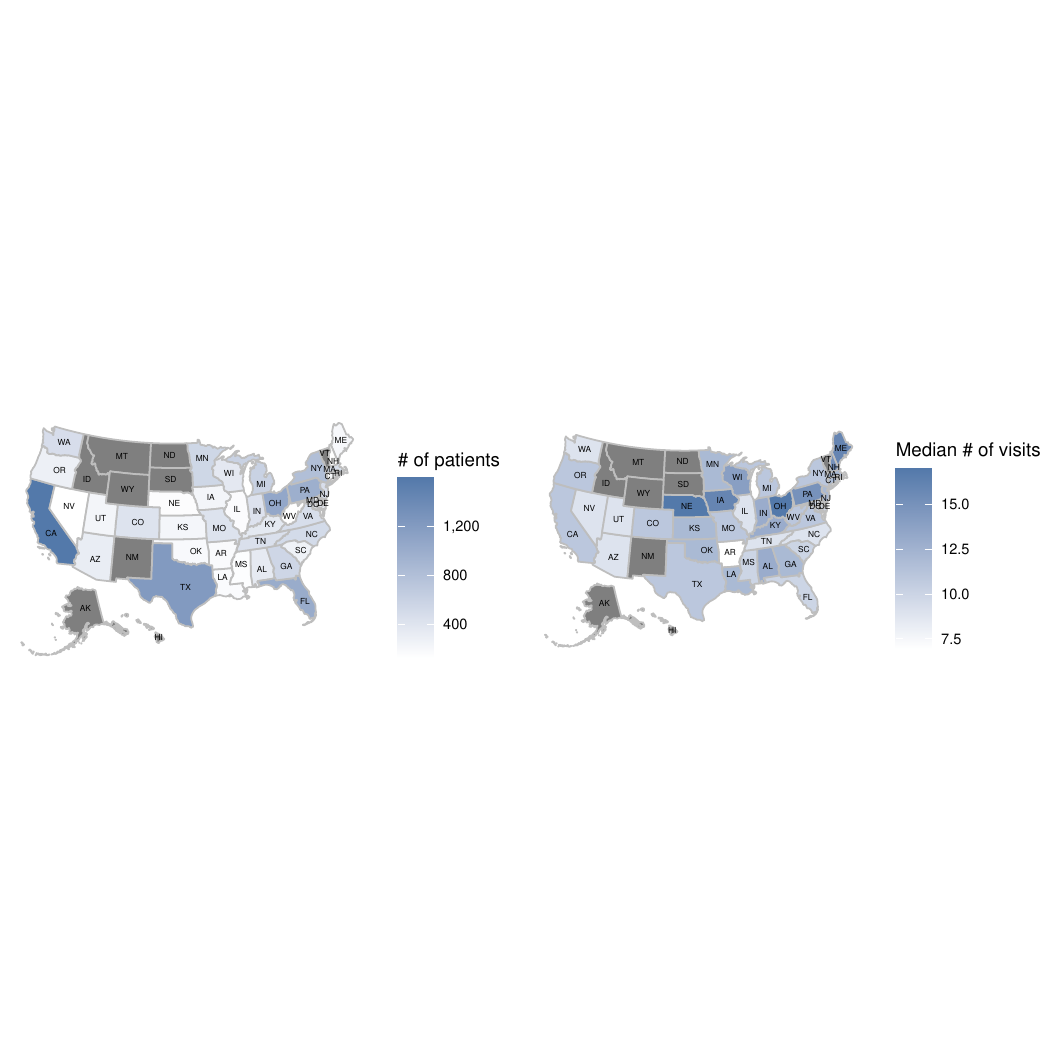}}
    \caption{The number of patients and the median number of visits per state are presented. Grey areas indicate the stages where the analysis was not performed due to the small number of patients (n $<$ 120).}
    \label{fig:Des1}
\end{figure}

\begin{figure}[!ht]
    \centerline{%
    \includegraphics[width = 15cm]{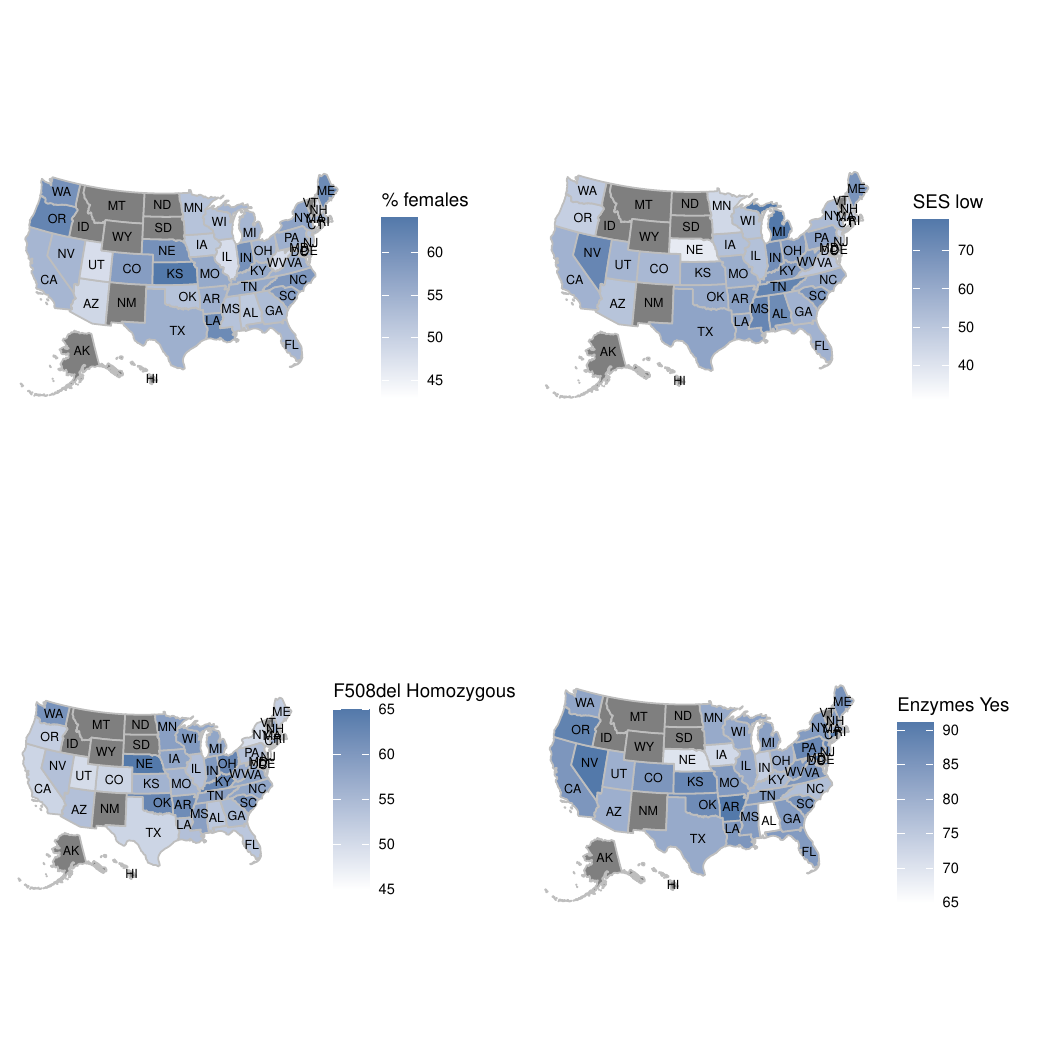}}
    \caption{Descriptive statistics for the variables included in the model per state. These variables include gender, socioeconomic status, F508del homozygous, and pancreatic enzyme use (from top-left to bottom-right).}
    \label{fig:Des2}
\end{figure}

\begin{figure}[!ht]
    \centerline{%
    \includegraphics[width = 12cm]{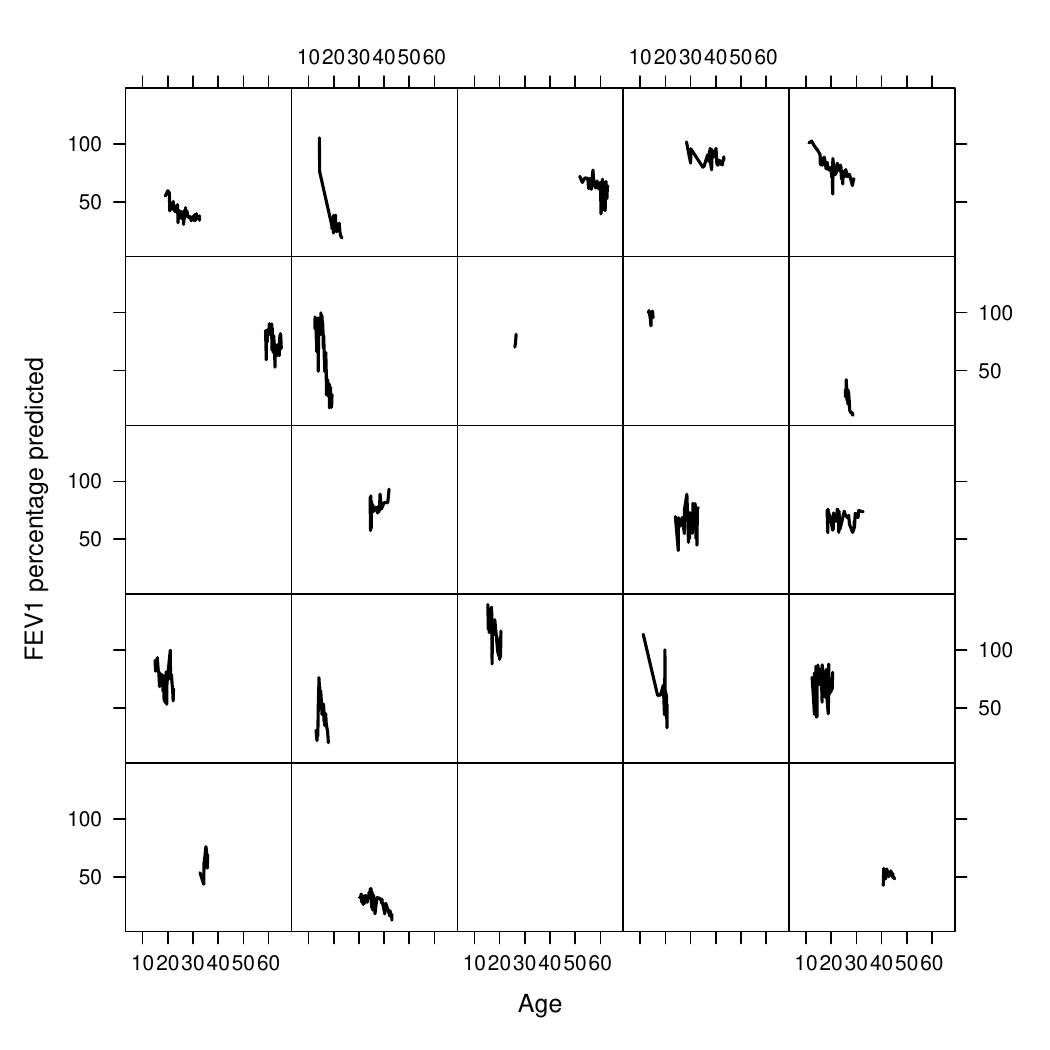}}
    \caption{Example of associations between encounter age (x-axis) and FEV$_1$ percentage predicted (y-axis) for various CF subjects.}
    \label{fig:des_FEV1}
\end{figure}

\begin{figure}[!ht]
    \centerline{%
    \includegraphics[width = 12cm]{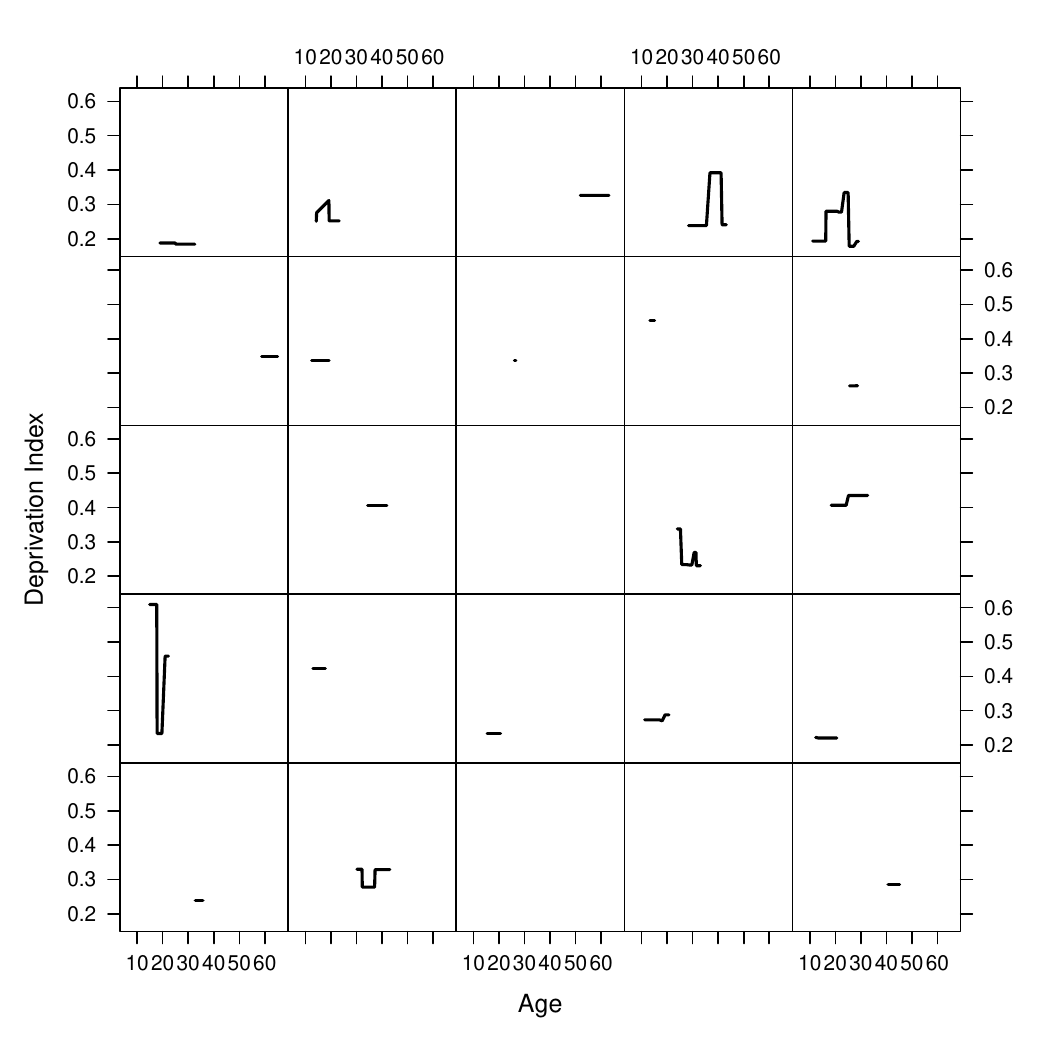}}
    \caption{Example of associations between different encounter age (x-axis) and deprivation index (y-axis) for various CF subjects.}
    \label{fig:des_depIndex}
\end{figure}

\begin{figure}[!ht]
    \centerline{%
    \includegraphics[width = 15cm]{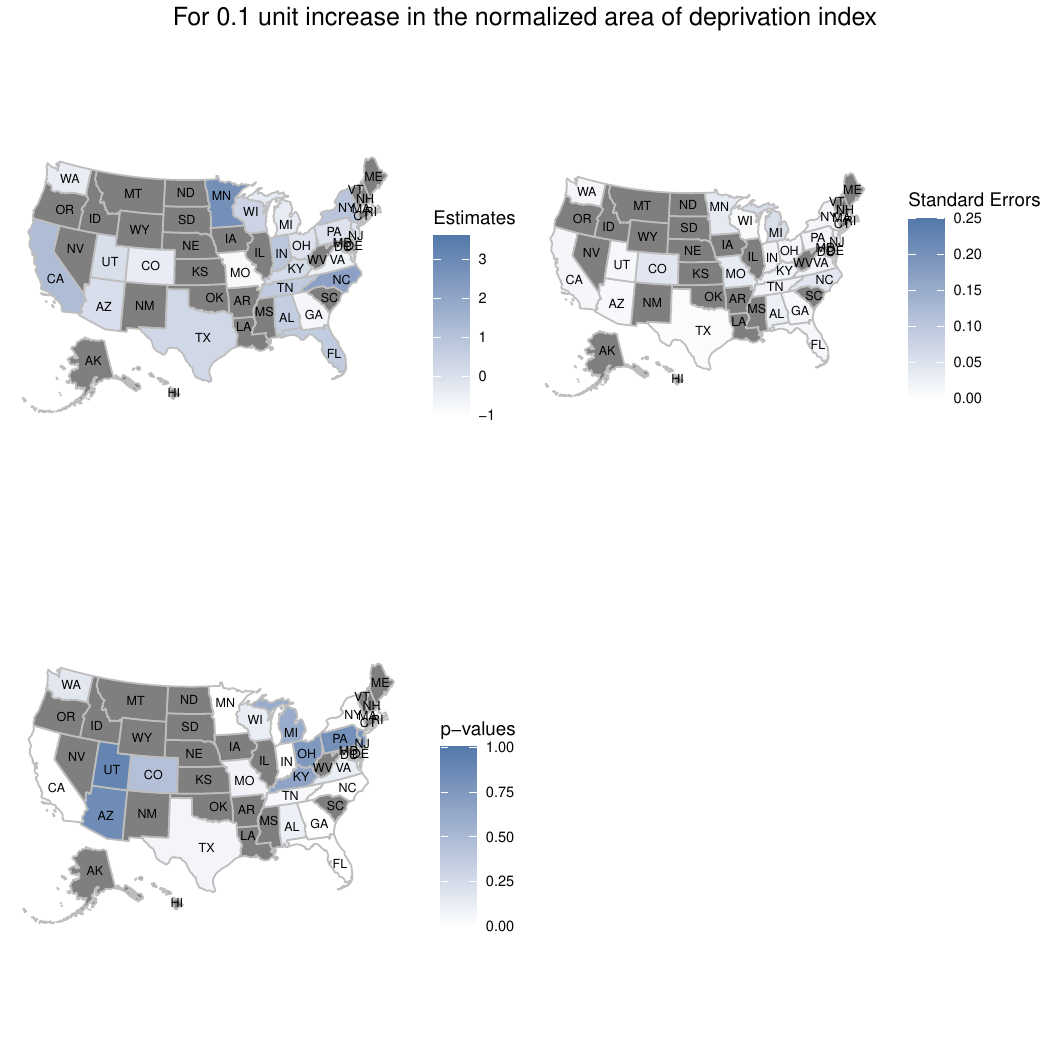}}
    \caption{Association parameters per state. Grey areas indicate the stages where the analysis was not performed due to the small number of patients (n $<$ 120, age $\leq$ 18). All longitudinal measurements were used.}
    \label{fig:res_all_18}
\end{figure}

\begin{figure}[!ht]
    \centerline{%
    \includegraphics[width = 15cm]{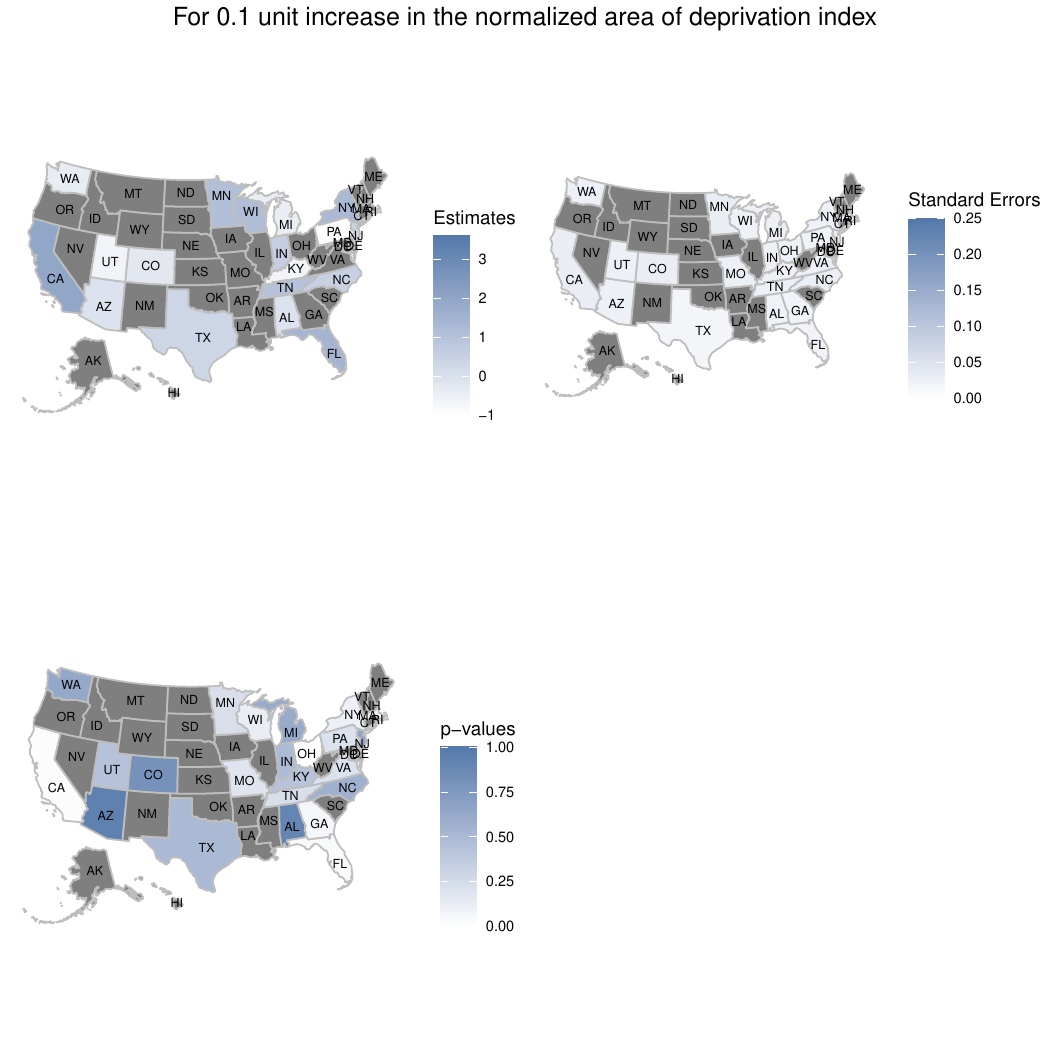}}
    \caption{Association parameters per state. Grey areas indicate the stages where the analysis was not performed due to the small number of patients (n $<$ 120, age $\leq$ 18). Two years of longitudinal measurements were used.}
    \label{fig:res_2years_18}
\end{figure}

\begin{figure}[!ht]
    \centerline{%
    \includegraphics[width = 15cm]{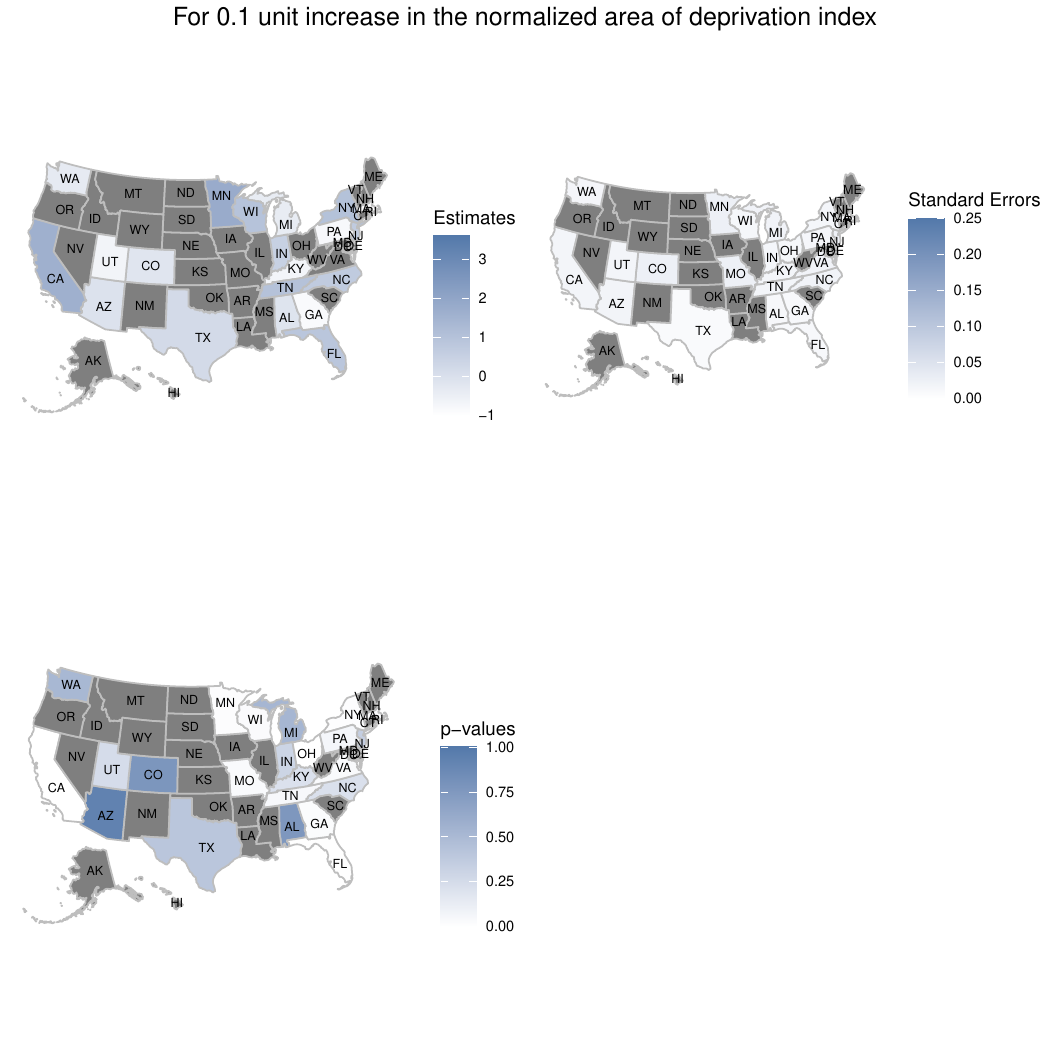}}
    \caption{Association parameters per state. Grey areas indicate the stages where the analysis was not performed due to the small number of patients (n $<$ 120, age $\leq$ 18). Five years of longitudinal measurements were used.}
    \label{fig:res_5years_18}
\end{figure}

\begin{figure}[!ht]
    \centerline{%
    \includegraphics[width = 15cm]{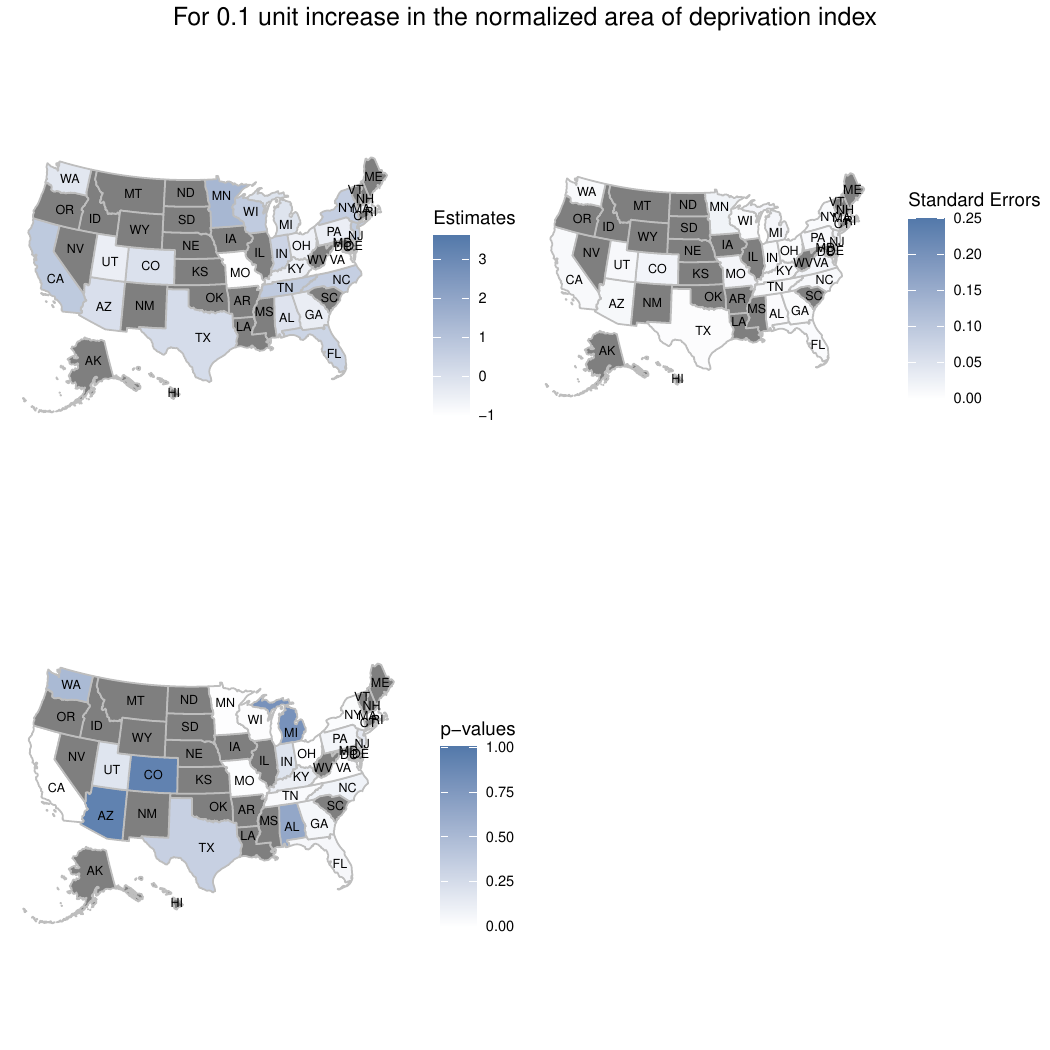}}
    \caption{Association parameters per state. Grey areas indicate the stages where the analysis was not performed due to the small number of patients (n $<$ 120, age $\leq$ 18). Ten years of longitudinal measurements were used.}
    \label{fig:res_10years_18}
\end{figure}

\begin{figure}[!ht]
    \centerline{%
    \includegraphics[width = 15cm]{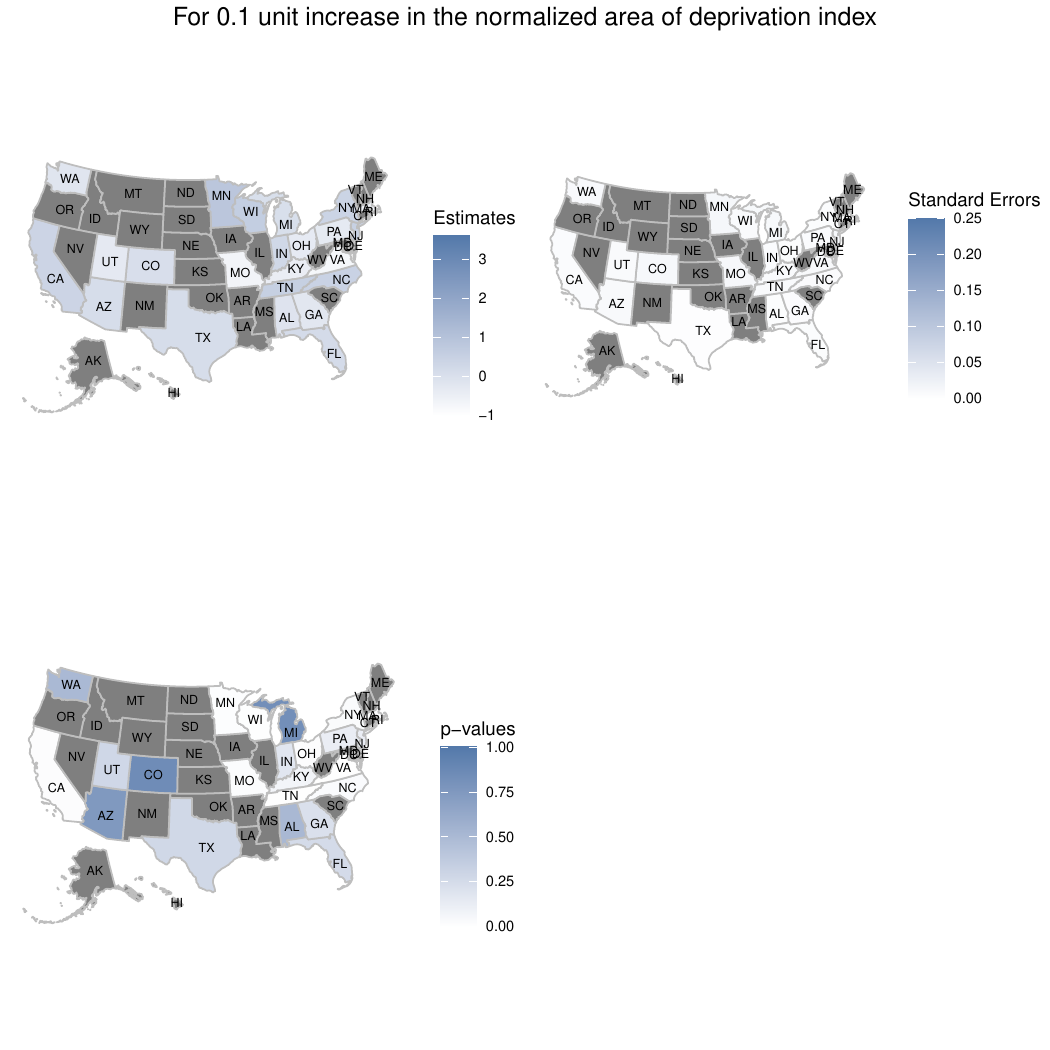}}
    \caption{Association parameters per state. Grey areas indicate the stages where the analysis was not performed due to the small number of patients (n $<$ 120, age $\leq$ 18). Fifteen years of longitudinal measurements were used.}
    \label{fig:res_15years_18}
\end{figure}

\begin{figure}[!ht]
    \centerline{%
    \includegraphics[width = 15cm]{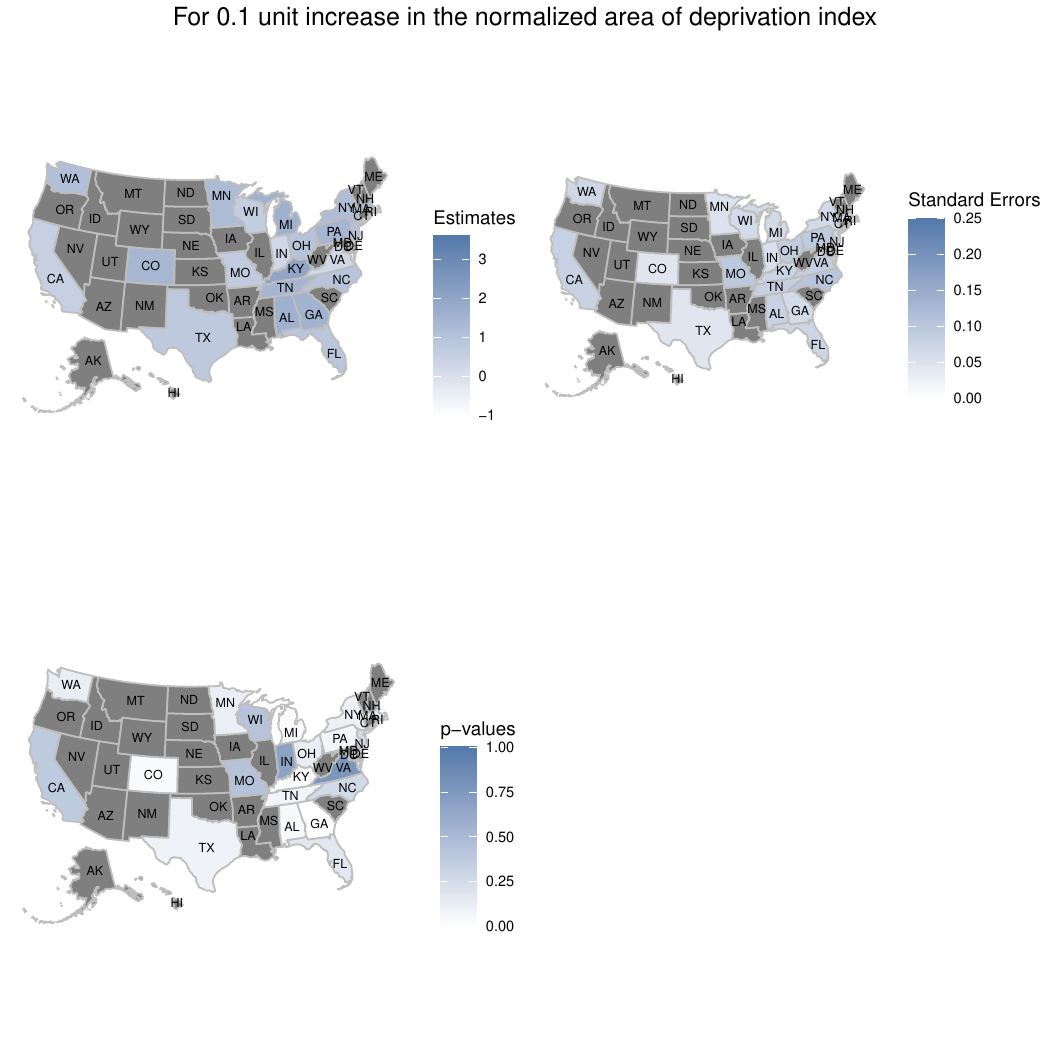}}
    \caption{Association parameters per state. Grey areas indicate the stages where the analysis was not performed due to the small number of patients (n $<$ 120, age $>$ 12). All longitudinal measurements were used.}
    \label{fig:res_all_12}
\end{figure}

\begin{figure}[!ht]
    \centerline{%
    \includegraphics[width = 15cm]{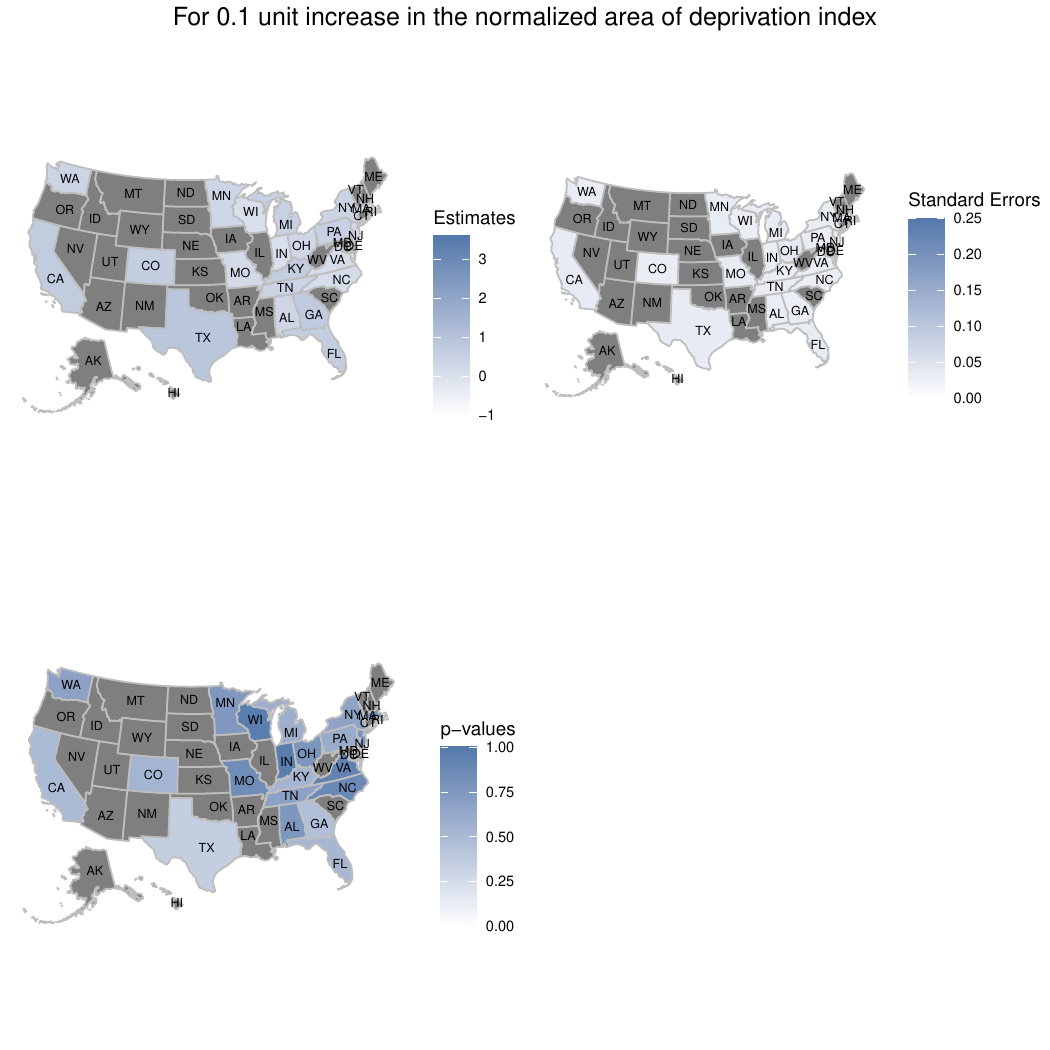}}
    \caption{Association parameters per state. Grey areas indicate the stages where the analysis was not performed due to the small number of patients (n $<$ 120, age $>$ 12). Two years of longitudinal measurements were used.}
    \label{fig:res_2years_12}
\end{figure}

\begin{figure}[!ht]
    \centerline{%
    \includegraphics[width = 15cm]{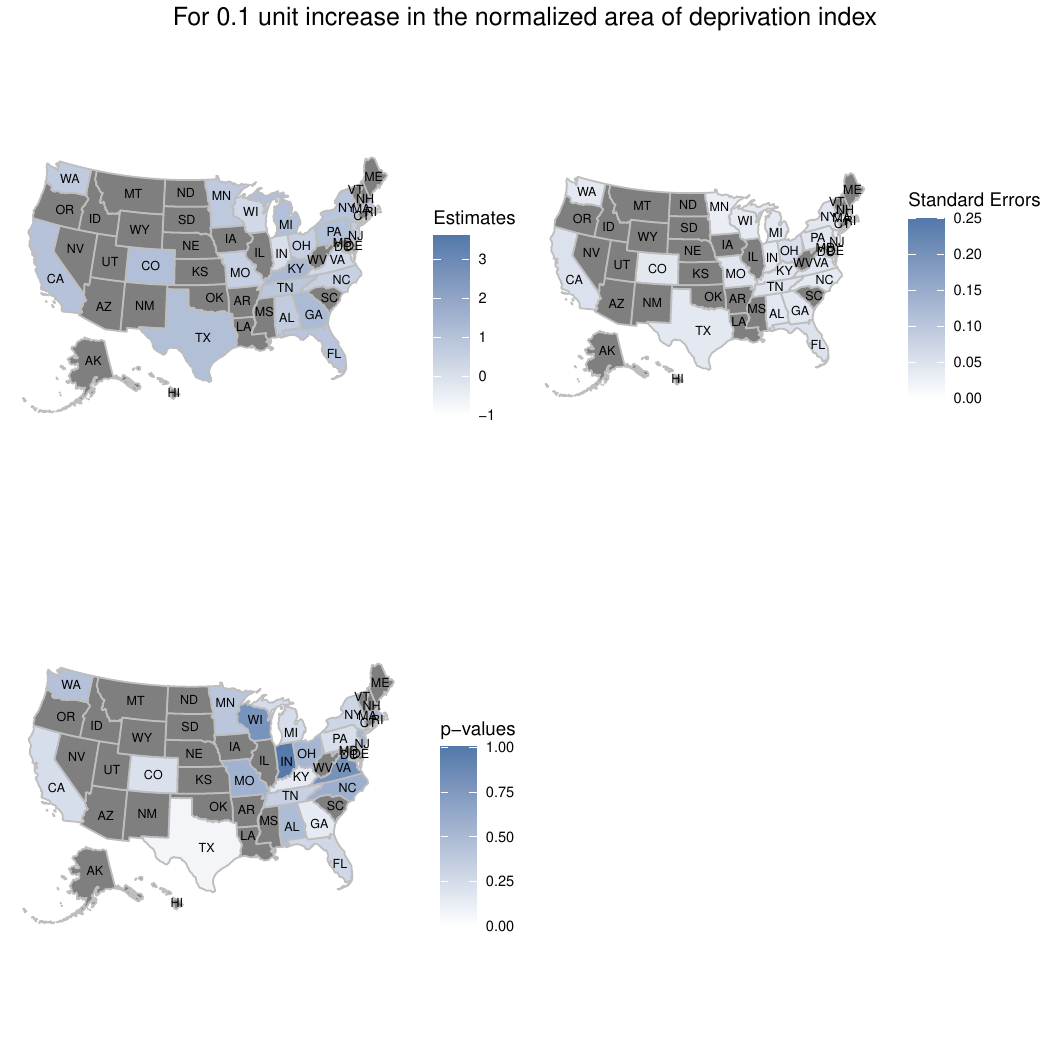}}
    \caption{Association parameters per state. Grey areas indicate the stages where the analysis was not performed due to the small number of patients (n $<$ 120, age $>$ 12). Five years of longitudinal measurements were used.}
    \label{fig:res_5years_12}
\end{figure}

\begin{figure}[!ht]
    \centerline{%
    \includegraphics[width = 15cm]{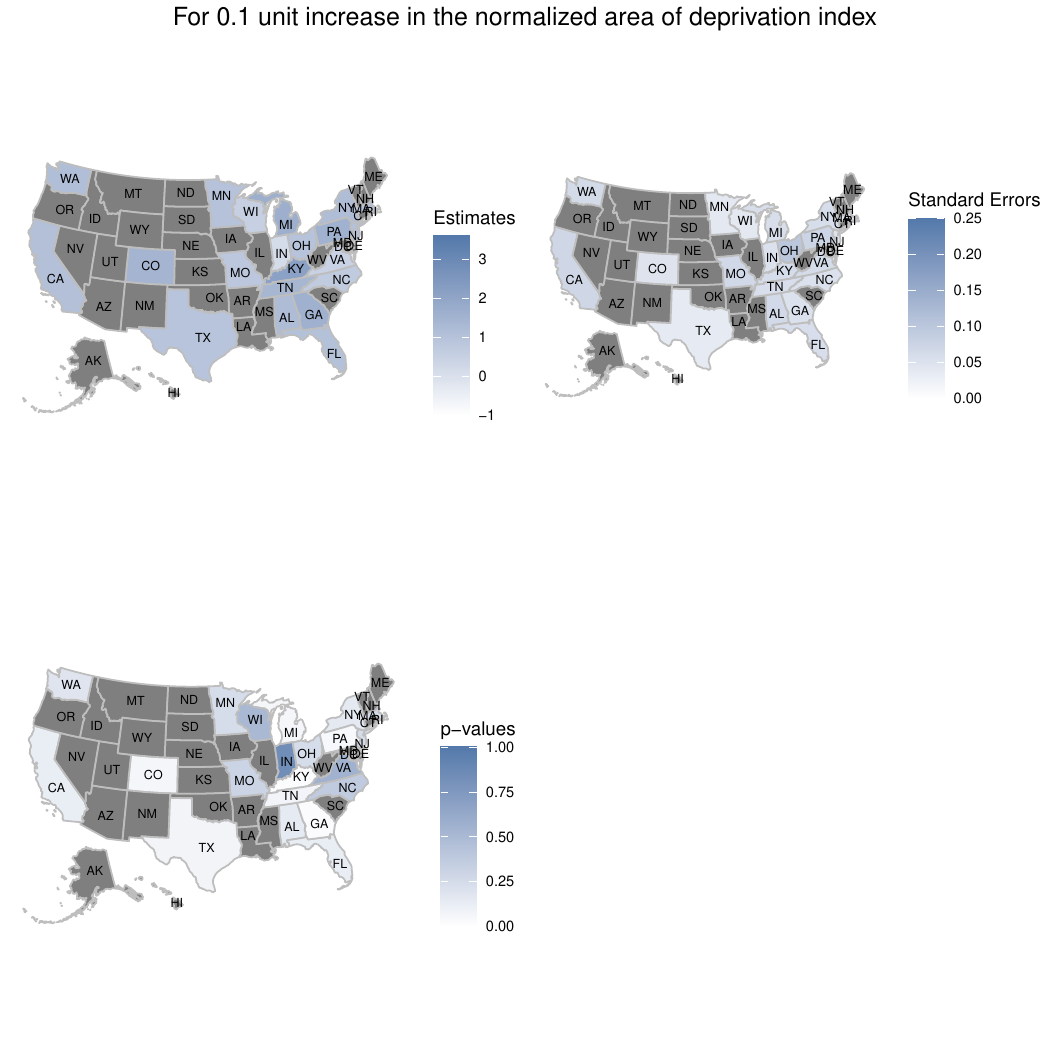}}
    \caption{Association parameters per state. Grey areas indicate the stages where the analysis was not performed due to the small number of patients (n $<$ 120, age $>$ 12). Ten years of longitudinal measurements were used.}
    \label{fig:res_10years_12}
\end{figure}

\begin{figure}[!ht]
    \centerline{%
    \includegraphics[width = 15cm]{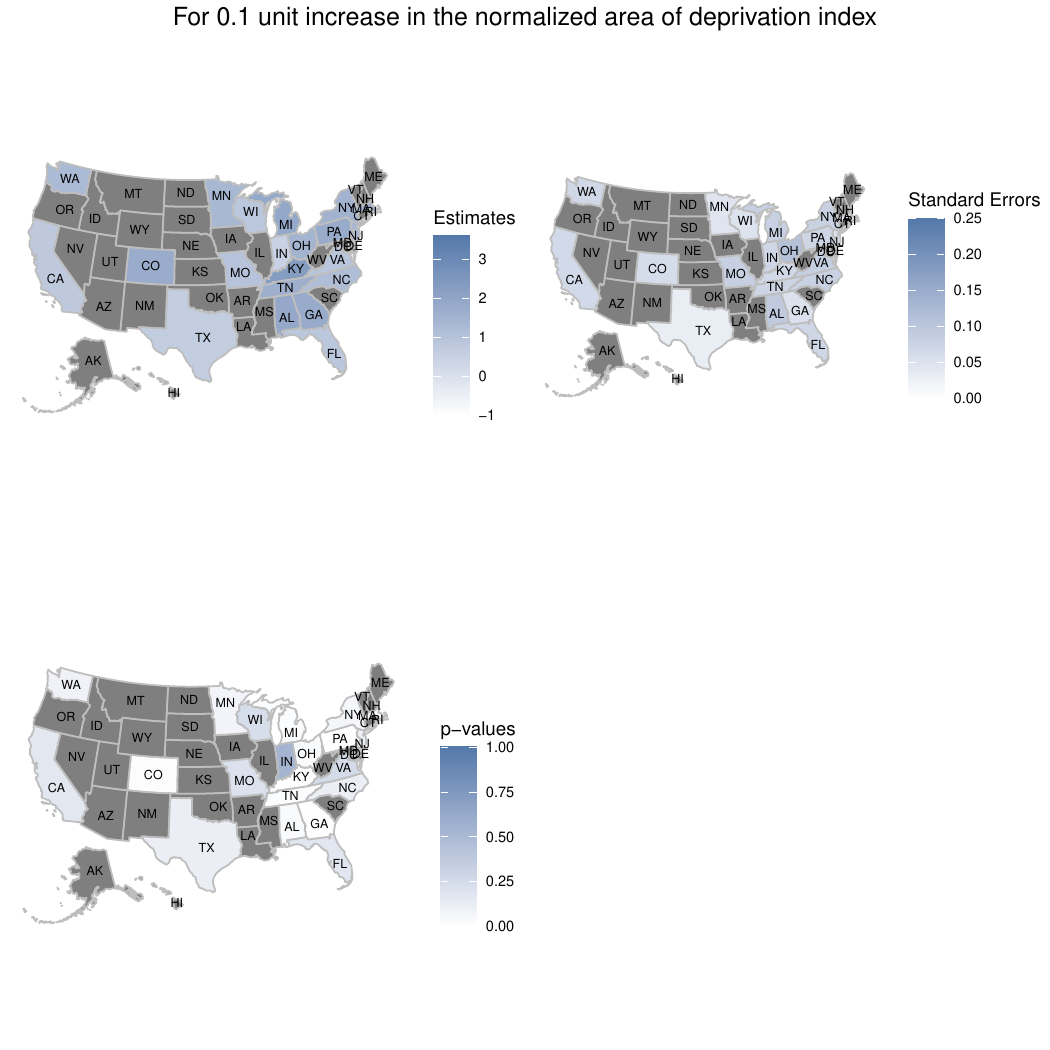}}
    \caption{Association parameters per state. Grey areas indicate the stages where the analysis was not performed due to the small number of patients (n $<$ 120, age $>$ 12). Fifteen years of longitudinal measurements were used.}
    \label{fig:res_15years_12}
\end{figure}

\clearpage

\end{document}